\documentclass[aps, prd, twocolumn, superscriptaddress, longbibliography, nofootinbib]{revtex4-2}
\usepackage[utf8]{inputenc}
\usepackage[dvipsnames]{xcolor}
\usepackage{graphicx}
\usepackage{bm}
\usepackage{amsmath}
\usepackage{siunitx}
\usepackage{hhline}
\usepackage{multirow}
\usepackage{tabularx}
\usepackage{comment}
\usepackage{amssymb} 
\usepackage{enumerate}
\usepackage{tensor}
\usepackage{mathrsfs}
\usepackage{hyperref}
\usepackage[normalem]{ulem}
\usepackage{accents}
\usepackage{scalerel}
\usepackage{braket}
\usepackage{mathtools} 

\makeatletter
\newcommand*{\leqdef}{\mathrel{\rlap{%
			\raisebox{0.25ex}{$\m@th\cdot$}}%
		\raisebox{-0.25ex}{$\m@th\cdot$}}%
	=}
\newcommand*{\reqdef}{=\mathrel{\rlap{%
			\raisebox{0.25ex}{$\m@th\cdot$}}%
		\raisebox{-0.25ex}{$\m@th\cdot$}}
}
\makeatother

\definecolor{DarkGreen}{rgb}{.1, .5, .1}

\begin{document}
	
	\title{Gravitational wave energy momentum-tensor in reduced Horndeski theories} 
	
	\author{João C. Lobato}
	
	\affiliation{Universidade Federal do Rio de Janeiro,
		Instituto de F\'\i sica\\
		CEP 21941-972 Rio de Janeiro, RJ, Brazil}
	
	\author{Maurício O. Calvão}

 	\affiliation{Universidade Federal do Rio de Janeiro,
		Instituto de F\'\i sica\\
		CEP 21941-972 Rio de Janeiro, RJ, Brazil}
	
	\begin{abstract}
	We generalize, imposing the field equations only at dominant order, the  Isaacson formula for the gravitational wave (GW) energy-momentum tensor  (EMT) to the class  of Horndeski theories in which the tensor modes travel at the speed of light (reduced Horndeski theories) and scalar waves are present. We discuss important particular cases such as: theories where scalar waves are also luminal and theories in which the transverse-traceless gauge can be achieved in an arbitrary open set. The  vanishing of the trace of the gravitational wave energy-momentum tensor is obtained for theories in which all wave perturbations propagate at the speed of light. The trace is shown not to vanish trivially in other cases. We obtain, as a particular case of our general result, the GW EMTs, in a Brans-Dicke theory, both in the Einstein frame, recovering previous results in the literature, and in the Jordan frame, thereby showing the GW EMT is not conformally invariant.  We further prove that there exists a subclass of reduced Horndeski theories where, in contrast to general relativity, the divergence of the GW EMT does not vanish even after the imposition of the full equations of motion, assuming an eikonal solution.
	\end{abstract}
	
	\maketitle

\section{Introduction}

The recent first detection of the stochastic gravitational wave background (SGWB) by NANOGrav \cite{Agazie2023I} has proven once more that there are still numerous new wonders yet to be discovered about our universe. It has been responsible for an additional exciting chapter in the continuous fresh series of discoveries surrounding gravitational wave (GW) physics and its power to give us new physics information will be enormous \cite{Agazie2023II, Afzal2023}. The first space based interferometer to be launched in the next decade, the Laser Interferometer Space Antenna (LISA) \cite{Amaro2017, Baker2020}, will be able to provide another angle to the vast GW landscape, improving our understanding of fundamental physics, astrophysics and cosmology \cite{Barausse2020, Amaro2023, Auclair2023}. In particular, the SGWB coming from compact white dwarf binaries in our own galaxy is expected to be detected by LISA \cite{Romano2017}.

The SGWB will serve as a new probe for alternative theories of gravity as well. Short duration GWs need a number of interferometers in different parts of the globe if the goal is to measure additional polarizations predicted by  modified theories of gravity. But, with the SGWB continuous signal, it is possible to detect such polarizations with a single detector provided a  sufficiently long observation time is employed \cite{Abbot2018}. 

Another way SGWB can indicate deviations from GR is if direct corrections to the detected signal exist. As pointed out by \cite{Isi2018, Lobato2022}, usually when discussing the SGWB at a theoretical level, one pays attention to the spectral energy density per unit solid angle but takes for granted the way this physical quantity relates to the actually measurable SGWB signal. This relation between what is directly measured and what is more physically relevant can change in modified gravity. In \cite{Isi2018}, this relation is generalized to Brans-Dicke,  Chern-Simons and massive gravity theories. In \cite{Lobato2022}, we obtained the aforementioned relation, assuming there are no scalar waves, in the context of reduced Horndeski theories, the most general scalar-tensor theory having second-order equations of motion originating from a non-degenerate Lagrangian with tensor waves traveling at the speed of light. The luminal property of tensorial perturbations is observably justified, at least for low redshifts, by multimessenger sources \cite{Abbot2017, Goldstein2017}. 

The equation relating the directly measurable SGWB signal and the spectral energy density of GWs has corrections for alternate theories of gravity because, among other possible causes such as a change in GW propagation law, the energy-momentum  tensor (EMT) of GWs changes when compared to its simple form in general relativity (GR). The GW EMT arises from the non-linear character of the field equations. When assuming the presence of high-frequency GW perturbations traveling through the background space-time, low frequency contributions related with the GWs arise in the background part of the field equations, a back-reaction mechanism that describes how GWs, while being a result of the matter curved space-time, can themselves generate more curvature. This process was first studied in \cite{Isaacson1968I, Isaacson1968II} in the GR context and further mathematically developed by \cite{Burnett1989, Green2011, Green2013}.

In \cite{Stein2011}, many crucial aspects regarding the computation of the GW EMT are revised and generalized and the tensor is obtained for some alternate gravity theories, such as Chern-Simons and scalar-tensor theories where the scalar field has a canonical kinetic term in the action together with an interaction term coupling the scalar field and scalar curvature invariants of second rank or higher, that is, those constructed with the contraction of at least three curvature related tensors. Such a family of theories does not include the Lagrangian term  $f(\varphi) R$, where $\varphi$ is the dynamical scalar field, where $R$ is the Ricci scalar. Another particularity of the study made in \cite{Stein2011} is that only asymptotically flat space-times are taken into account. Thus,  important systems, such as cosmological  ones, cannot be included in those results.  

In \cite{Preston2014, Preston2016} the GW EMT is obtained in $f(R)$ theories. In GR, the tensor has always vanishing trace, being then interpreted as describing a radiation fluid, while in $f(R)$ it can have negative values, which allows the resultant fluid to be understood, in principle, as dark energy, responsible for the present accelerated expansion stage of the universe. 

The GW EMT for reduced Horndeski theories was originally obtained in \cite{Dalang2020}. But in this work the presence of scalar waves was completely neglected. Further, several results regarding scalar waves were derived in \cite{Dalang2021}, but the GW EMT was not generalized to contain scalar wave contributions. 

In this present paper, our main goal is to obtain the off-shell GW EMT in the context of reduced Horndeski theories assuming scalar waves. Several new features must be considered: in this context, the transverse-traceless (TT) gauge cannot always be achieved, scalar waves do not always travel at the speed of light, the true degrees of freedom of GWs are mixed together inside the tensor perturbations of the metric and some properties of the weak-limit average used to obtain the tensor have to be reviewed.

We do not impose the full field equation in the most general context, but some theories are explored in which, even after the field equations are used, the divergence of the GW EMT is shown not to vanish. This implies that the duality relation for the GW analogous cosmological distances (the luminosity and angular diameter distances) is altered, which, as will be discussed further in the text, can imply modifications on the spectral energy density of the SGWB. 

Some particular subcases of interest are explored such as Horndeski theories with luminal scalar waves and those in which the TT gauge can be achieved. In the first case, the trace of the GW EMT is shown to vanish, as one should expect of a radiation fluid, while in the other it is shown not to vanish in a trivial way, which could potentially indicate a dark energy fluid, just as suggested in \cite{Preston2016} for $f(R)$ gravity.  

In summary, the most important results of our study are (see Fig.~(\ref{summary})):
\begin{itemize}
	\item We obtain the GW EMT for a general reduced Horndeski theory, but only imposing the equations of motion at dominant order (Eqs.~(\ref{complete_EMT}), (\ref{T_2}), (\ref{T_4})), (\ref{T_3}).
	\item We obtain the GW EMT for reduced Horndeski theories where scalar waves are luminal and show that its trace vanishes (Eqs.~(\ref{luminal_GW_EMT}) and (\ref{vanishing_trace})).
	\item We obtain the GW EMT for reduced Horndeski theories where the transverse-traceless gauge is achievable, showing that its trace does not vanish trivially and thus may indicate an alternative for dark energy (Eqs.(\ref{TT_EMT}) and (\ref{TT_gauge})).
	\item We reobtain the GW EMT for Brans-Dicke theory in the Einstein frame and confirm previous results from literature (Eq.~(\ref{Jordan_EMT})).
	\item We obtain the GW EMT for Brans-Dicke theory in the Jordan frame and show that this object is not conformally invariant (Eqs.~(\ref{EMT_Einstein_frame}), (\ref{Einstein_EMT}) and (\ref{Jordan_to_Einstein_EMT})).
	\item We give an example of a reduced Horndeski theory where, even after imposing the complete field equations and assuming an eikonal solution, the divergence of the GW EMT does not vanish, which impacts the GW duality relation and the SGWB signal (Eqs.(\ref{particular_theory}), (\ref{particular_EMT}) and (\ref{vanishing_divergence})).
\end{itemize}

The results here can originate further investigation of possible modifications of gravity that can be tested by several different ways, including the SGWB signal analysis.

In Section \ref{sec:preliminaries} we present the theory we work on, define the GW EMT in a general context, describe tensor and scalar wave propagation, discuss some limitations on fixing gauges and present the weak-limit average. In Section \ref{sec:expansion} we expand the action until second order in the amplitude of GWs. In Section \ref{sec:EMT} we vary the action and take the weak limit average to obtain the GW EMT. In Section \ref{sec:special cases} we further explore particular cases of interest and reobtain the tensor in the Brans-Dicke theory, confirming what was already known from literature. In Section \ref{sec: divergence} we calculate the divergence of the GW EMT in a particular subclass of Horndeski and prove that, even after the imposition of the field equation, it does not vanish for a general eikonal wave solution.

We use units in which $c=1$ and a Lorentzian metric with signature $+2$. The Riemann will be defined by satisfying 
\begin{align}
    v_{\alpha;\beta \gamma} - v_{\alpha;\gamma \beta} \reqdef {R^{\lambda}}_{\alpha \beta \gamma} v_{\lambda}
\end{align}
where $v_{\alpha}$ is a generic covector and we define the Ricci tensor as:
\begin{align}
    R_{\alpha \beta} \leqdef {R^{\lambda}}_{\alpha \lambda \beta}.
\end{align}

\section{Preliminaries} \label{sec:preliminaries}

\subsection{Reduced Horndeski theories}

The reduced Horndeski theory gravitational action \footnote{We will not consider matter action in this study.} is:
\begin{align}
	S_g = \frac{1}{16\pi G}\int\sqrt{-g} [G_2(\varphi,X) &+ G_3(\varphi,X) \Box \varphi \nonumber \\ &+ G_4(\varphi) R] d^4x, \label{Horndeski_action}
\end{align}
where $\varphi$ is a scalar field, $\Box \varphi \leqdef g^{\mu\nu} \varphi_{;\mu \nu}$ and $X \leqdef - \varphi^{,\mu}\varphi_{,\mu}/2$ is the kinetic energy of the scalar field. It is the most general non-degenerate scalar-tensor four-dimensional theory with second-order differential field equations in which the tensorial modes of GWs travel at the speed of light. Notice that no restriction is made for the speed of scalar waves. GR is the special case in which $G_2 = -2\Lambda$, $G_3 = 0 $ and $G_4 = 1$. \footnote{ For a comprehensive review on Horndeski theory, see \cite{Kobayashi2019}.}

The family of scalar-tensor theories in which the GW EMT was obtained in \cite{Stein2011} has the gravitational action: 
\begin{align}
    S_{g}' = S_{EH} + S_{int} + S_{\varphi}, \label{Steinaction}
\end{align}
where 
\begin{align}
    S_{EH} \leqdef \frac{1}{16 \pi G} \int \sqrt{-g} R d^4x
\end{align}
is the GR 
 gravitational action, 
\begin{align}
    S_{\varphi} \leqdef \beta \int \sqrt{-g} [X+ V(\varphi)] d^4x
\end{align}
is the canonical scalar term, with $\beta$ being a real constant and
\begin{align}
    S_{int} \leqdef \alpha \int \sqrt{-g} f(\varphi) \mathcal{R} d^4x, 
\end{align} 
where $\alpha$ is another real constant and $\mathcal{R}$ is a scalar built with curvature tensors (\textit{i.e.} Riemann and Ricci tensors.), provided $\mathcal{R}\neq R$.
Actions of Eqs~(\ref{Horndeski_action}) and (\ref{Steinaction}) do not coincide for at least two reasons. First, in Eq.~(\ref{Steinaction}), $G_2 = X + V(\varphi)$. Second, as stated before, the coupling between the scalar field and the curvature, present in $S_{int}$ is not of the form $G_4(\varphi) R$.

\subsection{Definition of GW EMT in scalar-tensor theories}

To obtain the GW EMT of a scalar-tensor theory, we first need to consider solutions of the field equations having gravitational waves traveling on top of a background spacetime with metric $\bar{g}_{\mu\nu}(x^\sigma)$. 

To that end, we will assume there is a coordinate system in which a two-parameter family of metrics satisfying the field equations can be split as:
\begin{align}
    g_{\mu \nu} (x^{\sigma}, \epsilon, \alpha) = \bar{g}_{\mu \nu}(x^{\sigma}) + \alpha h_{\mu \nu} (x^{\sigma}, \epsilon) \label{background_wave_split_2}
\end{align}
where $|\epsilon|, |\alpha| \ll 1$ are constant parameters (i.e. independent of $x^\sigma$) related to, respectively, the GW wavelength and amplitude ($h_{\mu \nu}$ is assumed to vary substantially more and to be extremely weaker than the background metric $\bar{g}_{\mu \nu}$). Henceforward, unless explicitly stated otherwise, all covariant derivatives and the raising and lowering of indices will be done with respect to the background metric. It is important to emphasize that the background metric in this work is generic, different from the treatment in \cite{Stein2011}, where $\bar{g}_{\mu \nu}$ is assumed to be asymptotically flat, not comprising important cases, such as the isotropic and homogeneous universes described by the Robertson-Walker metric.   

In the same spirit, one splits the scalar field as:
\begin{align}
    \varphi (x^{\mu}, \epsilon, \alpha) = \bar{\varphi}(x^{\mu}) + \alpha \delta \varphi (x^{\mu}, \epsilon). \label{background_wave_split}
\end{align}

Then, given the action of a theory, one can find the GW EMT by the following procedure. 

First, expand the action in powers of the GW amplitude:
\begin{align}
    S = \bar{S} + \alpha S^{(1)}(h_{\alpha \beta}, \delta \varphi) + \alpha^2 S^{(2)}(h_{\alpha \beta}, \delta \varphi) + \mathcal{O}(\alpha^3), \label{action_expansion}
\end{align}
where $h_{\mu \nu}$ and $\delta \varphi$ still depend on $\epsilon$, but, of course, $\bar{S}$ depends neither on $\epsilon$ nor on $\alpha$.

Second, vary the $\alpha$ dependent part of the action with respect to the background metric and take the weak limit average (which will be described further in the text, see Subsec.~\ref{subsec:weak_limit}): 
\begin{align}
    T^{(GW)}_{\mu \nu} \leqdef - \frac{2}{\sqrt{-\bar{g}}} \bigg\langle \frac{\delta (S-\bar{S})}{\delta \bar{g}^{\mu \nu}} \bigg\rangle \label{EMT_def}
\end{align}
As a good first approximation, our study will only be concerned with contributions coming from the linear and quadratic parts in Eq.~({\ref{action_expansion}}).

The most general treatment one can give for the GW EMT is to assume $\alpha$ and $\epsilon$ as independent. But, to simplify our discussion, we will assume from now on that these parameters  are equal:
\begin{align}
    \epsilon = \alpha. \label{amplitude_wavelength}
\end{align}
In principle, terms with different powers in $\epsilon$ can contribute to $S^{(1)}$ and $S^{(2)}$, depending on the action we are interested, since Eq.~(\ref{action_expansion}) is an expansion only in the $\alpha$ parameter. 

Notice that, in Eq.~(\ref{background_wave_split_2}), when we take the limit $\alpha \rightarrow 0$, the wave vanishes and the total metric reduces to the background one. But when Eq.~(\ref{amplitude_wavelength}) is assumed, it does not follow necessarily that
\begin{align}
    \lim_{\epsilon \rightarrow 0} \epsilon h_{\alpha \beta}(x^{\mu}, \epsilon) = 0.\label{vanishing_perturbation}
\end{align}
Here we will assume perturbations are such that Eq.~(\ref{vanishing_perturbation}) is valid. An example of physical significance  in which this is true is the eikonal function:
\begin{align}
    h_{\alpha \beta}(x^{\mu}, \epsilon) = H_{\alpha \beta}(x^{\mu}) e^{ik_{\nu}x^{\nu}/\epsilon}.
\end{align}
Similar behavior is assumed for the scalar perturbation.

\subsection{Gravitational waves in reduced Horndeski theories}

The evolution equation for the tensorial and scalar parts of gravitational wave can be compactly expressed as\footnote{For the complete expressions of each matrix element in terms of the Galileon functions, see \cite{Dalang2021}}:
\begin{align}
    &\bigg[\begin{pmatrix}
        K^{\alpha \beta} & K^{\rho \sigma \alpha \beta }\\ {K^{\alpha \beta}}_{\mu \nu} & {K^{\rho \sigma \alpha \beta }}_{\mu \nu} 
    \end{pmatrix} \nabla_{\alpha}\nabla_{\beta} + \begin{pmatrix}
        F^{\alpha} & F^{\rho \sigma \alpha }\\ {F^{\alpha }}_{\mu \nu} & {F^{\rho \sigma \alpha}}_{\mu \nu}
    \end{pmatrix} \nabla_{\alpha} 
    \nonumber \\&\hspace{80pt}+ \begin{pmatrix}
       M & M^{\rho \sigma}\\ M_{\mu \nu} & {M^{\rho \sigma }}_{\mu \nu}
    \end{pmatrix} \bigg] \begin{pmatrix}
        \delta \varphi \\ \hat{h}_{\rho \sigma}
    \end{pmatrix} = \begin{pmatrix}
        0 \\ 0
    \end{pmatrix},\label{coupled_GW_equations}
\end{align}
where $\hat{h}_{\alpha \beta}$ is the trace-reverse of $h_{\alpha \beta}$:
\begin{align}
    \hat{h}_{\alpha \beta} \leqdef h_{\alpha \beta} - \frac{h}{2}\bar{g}_{\alpha \beta} \label{trace_reverse}
\end{align}
and the matrices acting upon the perturbation fields are called kinetic, friction (or amplitude) and mass tensors.

In order to interpret the pair of perturbation fields in Eq.~(\ref{coupled_GW_equations}) as true degrees of freedom of GWs, the kinetic tensor acting upon the double covariant derivatives needs to be diagonal, as argued by \cite{Dalang2021}. In the most general case of a reduced Horndeski theory, this does not occur for the pair ($\delta \varphi, \hat{h}_{\alpha \beta}$). Nevertheless, the diagonalization of the kinetic term is still achievable if $\hat{h}_{\alpha \beta}$ is replaced by the new tensor \footnote{We choose here, differently to the notation adopted in \cite{Dalang2021}, to express $\hat{\gamma}_{\alpha \beta}$ with a hat since it is defined in terms of hatted quantities.}: 
\begin{align}
    \hat{\gamma}_{\alpha \beta} (x^{\mu}, \epsilon) \leqdef \hat{h}_{\alpha \beta}(x^{\mu}, \epsilon) + \hat{C}_{\alpha \beta}(x^{\mu}) \delta \varphi(x^{\mu}, \epsilon), \label{tensorial_dof}
\end{align}
where
\begin{align}
    \hat{C}_{\mu \nu} \leqdef \frac{1}{\bar{G}_{4}}(\bar{G}_{3,X}\bar{\varphi}_{,\mu}\bar{\varphi}_{,\nu} - \bar{G}_{4,\varphi}\bar{g}_{\mu \nu}),
\end{align}
and $\bar{G}_i$ stands for $G_i(\bar{\varphi}, \bar{X})$. The pair $(\delta \varphi, \hat{\gamma}_{\alpha \beta})$ can then be thought as the true degrees of freedom of GWs \cite{Dalang2021}. It is important to notice that the system of equations governing the evolution of this new pair is still coupled, only the kinetic term is diagonalized.

One can invert Eq.~(\ref{trace_reverse}), to give:
\begin{align}
	h_{\alpha \beta} = \hat{h}_{\alpha \beta} - \bar{g}_{\alpha \beta} \frac{\hat{h}}{2}. \label{h_in_terms_of_trace_reversed}
\end{align}
Then, substituting Eq.(\ref{tensorial_dof}) and its trace, one relates the initial perturbation with the real tensorial degree of freedom of the system:
\begin{align}
    h_{\alpha \beta} =  \hat{\gamma}_{\alpha \beta} -\frac{\bar{g}_{\alpha \beta}}{2}\hat{\gamma} - C_{\alpha \beta} \delta \varphi, \label{h_in_terms_of_gamma}
\end{align}
where 
\begin{align}
    	C_{\mu \nu} = \frac{1}{\bar{G}_{4}}[\bar{G}_{3,X}(\bar{\varphi}_{,\mu}\bar{\varphi}_{,\nu} + \bar{X} \bar{g}_{\mu \nu})+ \bar{G}_{4,\varphi}\bar{g}_{\mu \nu}] \label{C}
\end{align}
is the trace-reverse of $\hat{C}_{\mu \nu}$ \footnote{Notice that $\hat{\hat{A}}^{\alpha \beta} = A^{\alpha\beta}$, for any tensor $A^{\alpha \beta}$.}.

\subsection{Transverse-traceless gauge is not always achievable}

A very powerful and standard way of representing GWs in general relativity is by taking advantage of the gauge freedom the tensorial perturbation has to reduce the number of non-zero independent components to two: one related with the plus polarization and the other related with the cross polarization. The resulting tensor has trace zero and travels along null geodesics transversely with respect to the direction of propagation. The effect of GWs in free particles is then found to be of a shearing nature, that is, it is an anisotropic perturbation which preserves areas and volumes. The gauge in which such scheme is possible is called the transverse-traceless (TT) gauge. Despite its simplicity, the TT gauge is not always achievable in reduced Horndeski theories. 

Under the gauge change 
\begin{align}
h_{\mu \nu} \rightarrow h_{\mu \nu} + 2\xi_{(\mu;\nu)}, 
\end{align}
where $\xi_{\mu}$ is the gauge vector, the divergence of $\hat{\gamma}_{\mu \nu}$ transforms as
\begin{align}
    {\hat{\gamma}_{\mu \nu}}^{;\nu} \rightarrow {\hat{\gamma}_{\mu \nu}}^{;\nu} + \Box\xi_{\mu} + {\bar{R}^{\mu}}_{\nu}\xi^{\nu}.\end{align}
It is always possible to find a $\xi_{\mu}$ such that the harmonic gauge condition:
\begin{align}
{\hat{\gamma}_{\mu \nu}}^{;\nu} = 0 \label{harmonic_gauge}
\end{align}
is satisfied.

Although the harmonic gauge condition is necessary for obtaining the more restrictive TT gauge, it is not sufficient.  As argued in \cite{Dalang2020}, under the geometrical optics regime, the TT gauge can only be achieved along the whole GW null geodesic without violating Eq.~(\ref{harmonic_gauge}) if $G_{4,\varphi} = 0$ and $G_{3,X} = 0$. We will not restrict ourselves to this set of theories and will not adopt the TT gauge in the discussion that follows, although the harmonic gauge will be used throughout the study. This is a considerable difference when comparing our results with the usual procedure for obtaining the GW EMT in general relativity, since one expects, in the former case, the trace of the tensorial degree of freedom to be present.

\subsection{Unitary-harmonic gauge is not always achievable}

In \cite{Garoffolo2019}, a gauge in which Eq.~(\ref{harmonic_gauge}) is valid together with $\delta \varphi =0$ simultaneously is presented. This is called the harmonic-unitary gauge in \cite{Dalang2020}. In this gauge, all possible new contributions to the GW EMT we are trying to investigate would be only a gauge effect, not having physical consequences, since no scalar wave would result in real space-time curvature, not generating geodesic deviation between freely falling particles. 

But, as proven in \cite{Garoffolo2019}, this gauge is only achievable when scalar waves are luminal. And, even in this case, this is only a necessary condition. Since we are dealing with a family of theories in which scalar waves are not necessarily luminal and, even when they are luminal, this gauge cannot always be achieved, the calculation of the GW EMT in reduced Horndeski theories can still have physically  relevant new contributions.  

\subsection{The weak limit average} \label{subsec:weak_limit}

Here we discuss some features regarding the weak limit average \cite{Isaacson1968II, Burnett1989, Gravitation, Green2011, Stein2011, Preston2016} used to construct the GW EMT.

We will assume that although the perturbations have amplitudes of order $\epsilon$ (remembering Eq.~(\ref{amplitude_wavelength})), their derivatives can have  greater order. More specifically, we will impose that, for each derivative taken, the order of the resulting quantity will drop by a power of $\epsilon$:
\begin{align}
&\epsilon \hat{\gamma}_{\alpha \beta; \lambda_1 \lambda_2 ... \lambda_n} = \mathcal{O}(\epsilon^{1-n}),  \label{power_drop_tensor} \\
& \epsilon \delta \varphi_{; \lambda_1 \lambda_2 ... \lambda_n} = \mathcal{O}(\epsilon^{1-n}). \label{power_drop_scalar}
\end{align}
This rule allows us to inspect how GWs can curve the background space-time by contributing to the energy-momentum source of the background field equations. If we assumed that derivatives were of the same order as the perturbations, there would be no effect on the background space-time due to GWs. On the other hand, if we assumed the drop in powers of $\epsilon$ for each derivative taken to be greater than one, we would necessarily end up with divergent terms in the GW EMT.

The weak limit average of a given tensor $A_{\alpha_1 \alpha_2 ... \alpha_n} (x^{\mu}, \epsilon)$ is
\begin{align}
    \langle A_{\alpha_1 \alpha_2 ... \alpha_n} (x^{\mu}) \rangle \leqdef \lim_{\epsilon \rightarrow 0} \int \sqrt{-\bar{g}}&{f^{\beta_1...\beta_n}}_{\alpha_1 \alpha_2 ... \alpha_n}(x'^{\mu}, x^{\mu}) \nonumber \\ &\hspace{-25pt}\times A_{\beta_1 ... \beta_n}(x'^{\mu}, \epsilon) d^4x', \label{weak_limit}\end{align}
where $f^{\alpha_1 \alpha_2 ... \alpha_n}$ is a function that falls smoothly to zero for $|x^{\mu} - x'^{\mu}|$ of the order of several $\epsilon$ wavelengths but still small compared with the typical background scale of variation. The limit is made to enforce the smallness of the wavelength when compared with the scale in which the average is made.

Since $f^{\alpha_1 \alpha_2 ... \alpha_n}$ is negligible for background scales, background quantities, which do not depend on $\epsilon$ in any way, can be considered constant under the region of integration:
\begin{align}
    \langle \bar{B}_{\alpha_1 \alpha_2...\alpha_n} \rangle = \bar{B}_{\alpha_1 \alpha_2...\alpha_n}.
\end{align}
Furthermore, we assume the limits
\begin{align}
    &\lim_{\epsilon \rightarrow 0} \hat{\gamma}_{\alpha \beta}(x^{\mu}, \epsilon) = \hat{\gamma}_{\alpha \beta}(x^{\mu},0), \\
    &\lim_{\epsilon\rightarrow 0} \delta \varphi(x^{\mu}, \epsilon) = \delta \varphi(x^{\mu},0),
\end{align}
to be well defined, which allow us to conclude that
\begin{align}
    &\langle \epsilon \hat{\gamma}_{\alpha \beta}\rangle = 0, \\
    & \langle \epsilon \delta \varphi\rangle = 0.
\end{align}

Because of Eqs.~(\ref{power_drop_tensor}) and (\ref{power_drop_scalar}), first derivatives of perturbations are of background order. But one of the most important properties of the weak limit is that, since $f^{\alpha_1 \alpha_2 ... \alpha_n}$ does not depend on $\epsilon$, such derivatives average to zero. In the case of the $\hat{\gamma}_{\alpha \beta}$ perturbation:
\begin{align}
    \langle \epsilon \hat{\gamma}_{\alpha \beta; \lambda}\rangle &=  \lim_{\epsilon \rightarrow 0} \int \sqrt{-\bar{g}}{f^{\gamma \delta \phi}}_{\alpha \beta \lambda} \epsilon \hat{\gamma}_{\gamma \delta;\phi} d^4x' \nonumber \\ &= - \lim_{\epsilon \rightarrow 0} \int \sqrt{-\bar{g}}  {f^{\gamma \delta \phi}}_{\alpha \beta \lambda;\phi} \epsilon \hat{\gamma}_{\gamma \delta} d^4x' = 0. \label{derivatives_vanish}
\end{align}
By induction, the n-th order derivative of the perturbations vanishes as well:
\begin{align}
    \langle \epsilon \hat{\gamma}_{\alpha \beta; \lambda_1 \lambda_2 .. \lambda_n}\rangle = 0. \label{average_derivative_vanish}
\end{align}

Because of these properties, one can always use integration by parts in products of derivatives of perturbations. For example:
\begin{align}
    \langle \epsilon^2 \hat{\gamma}_{\alpha \beta; \lambda} \delta \varphi_{; \delta}\rangle &=  \langle \epsilon^2 (\hat{\gamma}_{\alpha \beta; \lambda} \delta \varphi)_{; \delta} \rangle - \langle \epsilon^2 \hat{\gamma}_{\alpha \beta; \lambda \delta} \delta \varphi\rangle \nonumber \\ &= - \langle \epsilon^2 \hat{\gamma}_{\alpha \beta; \lambda \delta} \delta \varphi\rangle, \label{by_parts}
\end{align}
since, analogously to Eq.~(\ref{derivatives_vanish}), one shows that $\langle \epsilon^2 (\hat{\gamma}_{\alpha \beta; \lambda} \delta \varphi)_{; \delta} \rangle = 0$. 

Another important property is that covariant derivatives of perturbations commute inside the average:
\begin{align}
    \langle \epsilon \hat{\gamma}_{\alpha \beta;\mu \nu}\rangle &= \langle \epsilon \hat{\gamma}_{\alpha \beta;\nu \mu} \rangle + {\bar{R}^{\tau}}_{\alpha \mu \nu}\langle\epsilon \hat{\gamma}_{\tau \beta} \rangle + {\bar{R}^{\tau}}_{\beta \mu \nu}\langle\epsilon \hat{\gamma}_{\alpha \tau} \rangle \nonumber \\ &= \langle \epsilon \hat{\gamma}_{\alpha \beta;\nu \mu} \rangle. \label{commutation}
\end{align}
Using Eqs.~(\ref{by_parts}) and (\ref{commutation}), we get:
\begin{align}
    \langle \epsilon^2 \hat{\gamma}_{\alpha \beta; \lambda} \delta \varphi_{; \delta}\rangle =  \langle \epsilon^2 \hat{\gamma}_{\alpha \beta; \delta} \delta \varphi_{; \lambda}\rangle. \label{derivative_exchange}
\end{align}

Having in mind the properties derived in this subsection, one can simplify the GW EMT defined in Eq.~(\ref{GW_EMT_definition}) \cite{Stein2011}. Terms appearing in $S^{(1)}$ will necessarily have a single perturbation or one of its derivatives. As a consequence of Eq.~(\ref{amplitude_wavelength}) and the property illustrated in Eq.~(\ref{average_derivative_vanish}), these terms will necessarily vanish under the average. This implies that we only need to compute $S^{(2)}$:
\begin{align}
    T^{(GW)}_{\mu \nu} \approx - \frac{2}{\sqrt{-\bar{g}}} \bigg\langle \epsilon^2 \frac{\delta S^{(2)}}{\delta \bar{g}^{\mu \nu}} \bigg\rangle. \label{GW_EMT_definition}
  \end{align}
This is the first approximation of the quantity related to GWs that will serve as an additional source to the background field equations of the theory.

\subsection{Gravitational wave energy momentum tensor}

In general relativity, one has the usual Isaacson formula:
\begin{align}
T^{GW}_{\mu \nu} = \frac{1}{32}\langle \epsilon^2 \hat{\gamma}_{\alpha \beta;\mu} {\hat{\gamma}^{\alpha \beta}}_{;\nu} \rangle = \frac{1}{32}\langle \epsilon^2 \hat{h}_{\alpha \beta;\mu} {\hat{h}^{\alpha \beta}}_{;\nu} \rangle.
\end{align}
where the last equality is valid because there is no scalar wave on the theory.

The correction to this tensor in reduced Horndeski theories was already obtained if we assumed scalar waves to be neglectible \cite{Dalang2020}:
\begin{align}
    T^{GW}_{\mu \nu} = \frac{\bar{G}_4}{32}\bigg\langle \epsilon^2 \bigg( \hat{h}_{\alpha \beta;\mu} {\hat{h}^{\alpha \beta}}_{;\nu} - \frac{1}{2}\hat{h}_{,\mu}\hat{h}_{,\nu}\bigg) \bigg\rangle,
\end{align}
where we cannot assume the second term to vanish since the TT gauge may not be achievable, as discussed previously.

Different corrections were obtained for the GW EMT when considering the presence of scalar waves for theories like Brans-Dicke, Chern-Simmons and massive gravity \cite{Isi2018, Stein2011}, but no article handled the whole family of reduced Horndeski theories with scalar waves. 

In this last case, one expects corrections to these expressions involving $\langle\delta \varphi_{,\mu} \delta \varphi_{,\nu}\rangle$, $\langle\delta \varphi_{,\mu} \hat{\gamma}_{,\nu}\rangle$, $\langle \hat{\gamma}_{,\mu} \hat{\gamma}_{,\nu}\rangle$ and $\langle S^{\alpha \beta} \hat{\gamma}_{\alpha \beta,\mu} \delta \varphi_{,\nu}\rangle$, with $S^{\alpha \beta}$ being a certain tensor to be deduced, where $\hat{h}_{\alpha \beta}$ is exchanged by $\hat{\gamma}_{\alpha \beta}$ because of the presence of the scalar waves. Since we do not assume a luminal nature for scalar perturbations, we still can have contributions related with $\Box{\delta \varphi}$ as well. Our aim is to express the GW EMT assuming all these new elements from the theory.

\section{Expanding the action} \label{sec:expansion}

We begin by expanding metric related quantities. Although the metric has only linear perturbations, the inverse metric needs to have a second order term in order to satisfy:
\begin{align}
    g_{\alpha \beta}g^{\beta \gamma} = {\delta^{\gamma}}_{\alpha},
\end{align}
to second order. This implies in the following expression, correct to second order:
\begin{align}
    g^{\mu \nu} = \bar{g}^{\mu \nu} - \epsilon h^{\mu \nu} + \epsilon^2 {h^{\mu}}_{\alpha} h^{\alpha \nu}. 
\end{align}
Here, one might ask why haven't we expanded the metric like
\begin{align}
    g_{\mu \nu} = \bar{g}_{\mu \nu} + \epsilon h_{\mu \nu}+ \epsilon^2 i_{\mu \nu}
\end{align}
since we are interested in contributions in the action until second order. We do not need to assume an $\epsilon^2$ term in the metric because it can only appear on $S^{ (2)}$ linearly, multiplying background terms.  Because of  Eq.~(\ref{average_derivative_vanish}), they would vanish in the weak limit average.

We now expand the metric determinant. Taking the determinant of Eq.~(\ref{background_wave_split}):
\begin{align}
    g = \bar{g} \det(I + \epsilon H), \label{metric_determinant}
\end{align}
where $g$ is the determinant of the total metric, $\bar{g}$ is the determinant of the background metric and ${(I + \epsilon H)^{\mu}}_{\nu} = {\delta^{\mu}}_{\nu}+\epsilon {h^{\mu}}_{\nu}$. Defining the logarithm of a matrix as:
\begin{align}
    \log (A) \leqdef \sum_{k=1}^{\infty} (-1)^{k+1} \frac{(A-I)^k}{k}, 
\end{align}
we may write, to second order in $\epsilon$, 
\begin{align}
    \det(I+\epsilon H) &= \exp[\log(\det(I+\epsilon H))] \nonumber \\ &= \exp[Tr(\log(I+ \epsilon H))] \nonumber \\ &= \exp\left(\epsilon Tr(H) - \frac{\epsilon^2}{2} Tr(H^2) \right) \nonumber \\ &= 1 + \epsilon Tr(H) + \frac{\epsilon^2}{2} \{[Tr(H)]^2 - Tr(H^2)\}, 
\end{align}
where in the second equality we have used a well known property (see \cite{MaggioreVol1}, for example) and $Tr$ is the trace of a matrix. Using this expansion in Eq.~(\ref{metric_determinant}):
\begin{align}
    \sqrt{-g} &=\sqrt{- \bar{g}} \left\{1 + \frac{\epsilon}{2}h + \frac{\epsilon^2}{8}\left[h^2 - 2{h^{\alpha}}_{\beta} {h^{\beta}}_{\alpha}\right]\right\},
\end{align}
where $h\leqdef Tr(H) = \bar{g}^{\alpha \beta}h_{\alpha \beta}$.

We next expand the scalar kinetic term $X$ using Eqs.~(\ref{background_wave_split_2}) and (\ref{background_wave_split}):
\begin{align}
    X = -\frac{1}{2} g^{\alpha \beta} \varphi_{;\alpha} \varphi_{;\beta} = \bar{X} + \epsilon X^{(1)} + \epsilon^2 X^{(2)}, \label{X_expansion}
\end{align}
where 
\begin{align}
    &\bar{X} = - \frac{1}{2}\bar{g}^{\alpha \beta} \bar{\varphi}_{,\alpha} \bar{\varphi}_{,\beta}, \label{X_bar} \\ &X^{(1)} = \frac{h^{\alpha \beta}}{2} \bar{\varphi}_{,\alpha} \bar{\varphi}_{,\beta} - \bar{g}^{\alpha \beta} \bar{\varphi}_{,\alpha} \delta \varphi_{,\beta}, \\
	&X^{(2)} = h^{\alpha \beta} \bar{\varphi}_{,\alpha} \delta \varphi_{,\beta} - \frac{\bar{g}^{\alpha \beta}}{2} \delta \varphi_{,\alpha} \delta \varphi_{,\beta} -\frac{1}{2} {h^{\alpha}}_{\mu} h^{\mu \beta} \bar{\varphi}_{,\alpha} \bar{\varphi}_{,\beta}.
\end{align}
Notice that, in Eq.~(\ref{X_expansion}), we have used the full metric to contract the indices instead of the background one, since we are interested in the action, which is expressed in terms of the complete metric. With this, the Galileon functions can be expanded:
\begin{align}
    G_i(\varphi, X) &= \bar{G}_i + \epsilon G_i^{(1)} + \epsilon^2 G_i^{(2)} \label{Galileon_expanded}
\end{align}
where
\begin{align}
    &G_i^{(1)} = \bar{G}_{i,\varphi} \delta \varphi + \bar{G}_{i,X} X^{(1)} \nonumber \\ &G_i^{(2)} =  \bar{G}_{i,X} X^{(2)} + \bar{G}_{i,\varphi \varphi} \frac{\delta \varphi^2}{2} + \bar{G}_{i,\varphi X} \delta \varphi X^{(1)} \nonumber \\ &\hspace{130pt}+ \bar{G}_{i,XX} \frac{{(X^{(1)})}^2}{2}. 
\end{align}

From the Ricci tensor expansion (see \cite{MaggioreVol1}, for example), one obtains the Ricci scalar to second order in $\epsilon$:
\begin{align}
    R = g^{\alpha \beta} R_{\alpha \beta} = \bar{R} + \epsilon R^{(1)} + \epsilon^2 R^{(2)}, \label{Ricci_expansion}
\end{align}
where
\begin{align}
    &R^{(1)} = {h_{\beta \alpha;}}^{\beta \alpha} - \bar{\Box} h - \bar{R}_{\mu \nu} h^{\mu\nu} \\ &R^{(2)} = \frac{1}{2} \bigg[\frac{3}{2} h_{\mu \alpha; \rho} {{h^{\mu \alpha}}_;}^{\rho} - h_{\mu \alpha; \rho} {{h^{\mu \rho}}_;}^{\alpha} \nonumber \\ &+ \left(\frac{h_{;\alpha}}{2} - {h_{\alpha \rho;}}^{\rho}\right)(2{h^{\alpha \mu}}_{;\mu} - {h_;}^{\alpha})\bigg] + {h^{\mu}}_{\alpha} h^{\alpha \nu} \bar{R}_{\mu \nu}\nonumber \\ &\hspace{65pt} + h^{\rho \alpha}(\bar{\Box} h_{\rho \alpha} + h_{;\rho \alpha} - {2{h^{\mu}}_{ \rho}}_{;(\mu\alpha)}),
\end{align}
and
\begin{align}
    {{h^{\mu}}_{ \rho}}_{;(\mu\alpha)} \leqdef \frac{1}{2}({{h^{\mu}}_{ \rho}}_{;\mu\alpha} + {{h^{\mu}}_{ \rho}}_{;\alpha\mu}).
\end{align}

To expand the D'alembertian in Eq.~(\ref{Horndeski_action}), we first need to expand the Christoffel coefficients:
\begin{align}
    \Gamma^{\alpha}_{\beta \gamma} = \bar{\Gamma}^{\alpha}_{\beta \gamma}+\epsilon\Gamma^{\alpha(1)}_{\beta \gamma} + \epsilon^2\Gamma^{\alpha(2)}_{\beta \gamma},
\end{align}
where
\begin{align}
    \Gamma^{\alpha(1)}_{\beta \gamma} = \frac{\bar{g}^{\alpha \lambda}}{2}(h_{\gamma \lambda;\beta} + h_{\beta \lambda;\gamma} - h_{\alpha \beta;\lambda})
\end{align}
and
\begin{align}
    \Gamma^{\alpha(2)}_{\beta \gamma} = - \frac{h^{\alpha \lambda}}{2}(h_{\gamma \lambda;\beta} + h_{\beta \lambda;\gamma} - h_{\alpha \beta;\lambda}).
\end{align}
Then, if $\nabla_{\alpha}$ is the covariant derivative with respect to $g_{\alpha \beta}$ we conclude 
\begin{align}
    \Box \varphi &= g^{\alpha \beta} \nabla_{\beta} \nabla_{\alpha}\varphi = g^{\alpha \beta} \nabla_{\beta}({\varphi_{,\alpha}})\nonumber \\ &= g^{\alpha \beta} [ \varphi_{,\alpha \beta} - (\bar{\Gamma}^{\gamma}_{\alpha \beta} + \epsilon \Gamma^{\gamma(1)}_{\alpha \beta}+\epsilon^2\Gamma^{\gamma(2)}_{\alpha \beta})\varphi_{,\gamma} ] \nonumber \\ &= \bar{\Box} \bar{\varphi} + \epsilon \bigg[\bar{\Box} \delta \varphi- \frac{\bar{g}^{\gamma \lambda}}{2}(2{h_{\mu \lambda}}^{;\mu} - h_{; \lambda})\bar{\varphi}_{,\gamma} - h^{\alpha \beta}\bar{\varphi}_{,\alpha \beta}\bigg] \nonumber \\ & \hspace{23pt}+ \epsilon^2\bigg[\bigg({h_{\mu \lambda}}^{;\mu} - \frac{h_{; \lambda}}{2}\bigg)\bigg(h^{\gamma \lambda}\bar{\varphi}_{,\gamma}-\bar{g}^{\gamma \lambda}\delta \varphi_{,\gamma}\bigg) \nonumber \\ & \hspace{50pt} + \frac{h^{\alpha \beta}\bar{g}^{\gamma \lambda}}{2}(2h_{\alpha \lambda;\beta} - h_{\alpha \beta; \lambda})\bar{\varphi}_{,\gamma} \nonumber \\ &\hspace{102pt}- h^{\alpha \beta} \delta \varphi_{,\alpha \beta}+{h^{\alpha}}_{\mu}h^{\mu \beta}\bar{\varphi}_{,\alpha \beta}\bigg].\label{Dalembertian}
\end{align}

Using Eqs.(\ref{X_expansion}), (\ref{Galileon_expanded}), (\ref{Dalembertian}) and (\ref{Ricci_expansion}), one finally finds
\begin{align}
    S^{(2)} = \sum_{i=2}^{4} S^{(2)}_i,
\end{align}
where
\begin{widetext}
\begin{align}
    S^{(2)}_{2} = \frac{\sqrt{-\bar{g}}}{16\pi G}\bigg\{&\frac{\bar{G}_2}{8}(h_{\alpha \beta}h^{\alpha \beta} - h^2) + \frac{h}{2}\bar{G}_{2,\varphi}\delta \varphi+ \frac{(\delta \varphi)^2}{2}\bar{G}_{2,\varphi\varphi} \nonumber \\&+ \bigg[\frac{1}{4} \bar{G}_{2,XX}[h_{\alpha \beta}\bar{\varphi}^{,\alpha} \bar{\varphi}^{,\beta} -2\bar{\varphi}^{,\alpha}\delta \varphi_{,\alpha}] +\delta \varphi\bar{G}_{2,\varphi X}\bigg] \frac{1}{2}(h_{\alpha \beta}\bar{\varphi}^{,\alpha} \bar{\varphi}^{,\beta} -2\bar{\varphi}^{,\alpha}\delta \varphi_{,\alpha})  \nonumber \\&+ \frac{1}{4}\bar{G}_{2,X}[-2\delta \varphi^{,\alpha}\delta \varphi_{,\alpha} - 2 h \delta \varphi^{,\alpha} \bar{\varphi}_{,\alpha} + \bar{\varphi}^{,\beta}(-2h^{\alpha \gamma}h_{\beta \gamma}\bar{\varphi}_{,\alpha} + h_{\alpha \beta}(4\delta \varphi^{,\alpha} + h \bar{\varphi}^{,\alpha} ))] \bigg\},  \label{S2_2}
\end{align}
\begin{align}
S^{(2)}_{3} = \frac{\sqrt{-\bar{g}}}{16\pi G}\bigg\{&\bar{G}_3\bigg[\frac{\bar{\varphi}^{,\alpha}}{4}[h(h_{,\alpha} - 2{h_{\alpha \gamma}}^{;\gamma})-2h^{\beta \gamma}(h_{\beta \gamma;\alpha}-2h_{\alpha \beta;\gamma}) -2 h_{\alpha \beta}(h^{,\beta} - 2{h^{\beta \gamma}}_{;\gamma})] \nonumber \\ &+ \frac{1}{8}[4h_{,\alpha}\delta \varphi^{,\alpha} - 8\delta \varphi_{,\alpha}{h^{\alpha \beta}}_{;\beta} -8h^{\alpha \beta}\delta \varphi_{;\alpha \beta} + 4 h \bar{\Box}\delta \varphi \nonumber \\ &+ \bar{\varphi}_{;\gamma \beta}(8{h_{\alpha}}^{\gamma}h^{\alpha \beta}-4hh^{\gamma \beta}) + \bar{\Box} \bar{\varphi}(h^2-2h_{\alpha \beta}h^{\alpha \beta})]\bigg] \nonumber \\ &+ \bar{G}_{3,\varphi}\frac{\delta \varphi}{2}[\bar{\varphi}_{,\alpha}(h^{,\alpha} - 2 {h^{\alpha \beta}}_{;\beta}) + 2\bar{\Box}\delta \varphi - 2 h^{\alpha \beta} \bar{\varphi}_{;\alpha \beta} + h \bar{\Box}\bar{\varphi} ] + \bar{G}_{3,\varphi \varphi}\frac{\delta \varphi^2}{2}\bar{\Box} \bar{\varphi}\nonumber \\ & + \bar{G}_{3,XX}\frac{\bar{\Box}\bar{\varphi}}{8}\bigg[h^{\alpha \beta} \bar{ \varphi}_{,\alpha} \bar{ \varphi}_{,\beta} -2 \bar{ \varphi}_{,\gamma}\delta \varphi^{,\gamma} \bigg]^2  +\bar{G}_{3,\varphi X}\bar{\Box}\bar{\varphi}\frac{\delta \varphi}{2} [h^{\alpha \beta}\bar{\varphi}_{,\alpha}\bar{\varphi}_{,\beta} - 2 \bar{\varphi}^{,\alpha} \delta \bar{\varphi}_{,\alpha}] \nonumber \\ &+\bar{G}_{3,X}\bigg[-\frac{\delta \varphi_{,\alpha}}{2}[(\delta \varphi^{,\alpha} - 2 h^{\alpha \beta}\bar{\varphi}_{,\beta})\bar{\Box}\bar{\varphi} + \bar{\varphi}^{,\alpha}(2\bar{\Box}\delta \varphi -2h^{\beta \gamma}\bar{ \varphi}_{;\beta \gamma}+h\bar{\Box}\bar{\varphi}) ] \nonumber \\ & + \frac{1}{4} \bar{\varphi}_{,\alpha}\bar{\varphi}_{,\beta}[h^{\alpha \beta}(2\bar{\Box} \delta \varphi + \bar{\varphi}_{,\gamma}(h^{,\gamma}-2{h^{\gamma \delta}}_{;\delta}) - 2 h^{\gamma \delta}\bar{\varphi}_{;\gamma \delta}+h\bar{\Box}\bar{\varphi})-2h^{\alpha \gamma}{h^{\beta}}_{ \gamma}\bar{\Box}\bar{\varphi} \nonumber \\ & \hspace{230pt}-2\delta \varphi^{,\alpha}(h^{,\beta} - 2{h^{\beta \gamma}}_{;\gamma})] \bigg]  \bigg\}, \label{S2_3}
\end{align}
\begin{align}
S^{(2)}_{4} = \frac{\sqrt{-\bar{g}}}{128\pi G}\bigg\{&\bar{G}_4\bigg[8{h_{\alpha}}^{\gamma}h^{\alpha \beta}\bar{R}_{\beta \gamma} - 4hh^{\beta \gamma}\bar{R}_{\beta \gamma} -2 h^{\alpha \beta}h_{\alpha \beta}\bar{R} + \bar{R}h^2 - 2h_{,\beta}h^{,\beta}\nonumber \\ &  - 8({h^{\alpha \beta}}_{;\alpha}-h^{,\beta}){h_{\beta \gamma}}^{;\gamma}+8h^{\alpha \beta}(h_{;\alpha \beta} -{{h_{\alpha \gamma}}^{;\gamma}}_{\beta}-{{h_{\alpha \gamma}}_{;\beta}}^{\gamma}+\bar{\Box}h_{\alpha \beta})\nonumber \\ & +4h[{h^{\beta \gamma}}_{;\beta \gamma} -\bar{\Box}h] + 2[3h_{\alpha \beta;\gamma}-2h_{\alpha \gamma;\beta}]h^{\alpha \beta;\gamma}\bigg]\nonumber \\ &+ 4 \bar{G}_{4,\varphi}\delta \varphi[h\bar{R}-2h^{\alpha \beta}\bar{R}_{\alpha \beta} + 2(h^{\alpha \beta}_{;\alpha \beta} - \bar{\Box}h)]+4\bar{G}_{4,\varphi \varphi}\bar{R}(\delta \varphi)^2\bigg\} \label{S^2_4}
\end{align}
\end{widetext}

We can further simplify Eqs.~(\ref{S2_3}) and (\ref{S^2_4}) by using integration by parts to express terms with two derivative acting upon a single perturbation as a product of first derivatives of perturbations. As an example:
\begin{align}
    4\bar{G}_4 h {h^{\beta \gamma}}_{;\beta \gamma} &=  -4[ \bar{G}_{4,\varphi} \bar{\varphi}_{,\gamma} h {h^{\beta \gamma}}_{;\beta} + \bar{G}_4 h_{,\gamma} {h^{\beta \gamma}}_{;\beta} ],
\end{align}
where the the total derivative term vanished, since fields are assumed to be zero at infinity, as usual. When considering the additional powers of $\epsilon$ coming from the derivative of perturbations (as in Eq.~(\ref{power_drop_tensor})), the first term in the above equation is of order $\epsilon$. It, then, vanishes in the weak limit average (variation with respect to background metric does not change the order of the term). This implies that one can make integration by parts only paying attention to the perturbation factors of each term. In the previous example:
\begin{align}
    4\bar{G}_4 h {h^{\beta \gamma}}_{;\beta \gamma} &=  -4\bar{G}_4 h_{,\gamma} {h^{\beta \gamma}}_{;\beta}. 
\end{align}

With this result, one simplifies:
\begin{widetext}
\begin{align}
    S^{(2)}_{3} = \frac{\sqrt{-\bar{g}}}{16\pi G}\bigg\{&\bar{G}_3\bigg[\frac{\bar{\varphi}^{,\alpha}}{4}[h(h_{,\alpha} - 2{h_{\alpha \gamma}}^{;\gamma})-2h^{\beta \gamma}(h_{\beta \gamma;\alpha}-2h_{\alpha \beta;\gamma}) -2 h_{\alpha \beta}(h^{,\beta} - 2{h^{\beta \gamma}}_{;\gamma})] \nonumber \\ &+ \frac{1}{8}[\bar{\varphi}_{;\gamma \beta}(8{h_{\alpha}}^{\gamma}h^{\alpha \beta}-4hh^{\gamma \beta}) + \bar{\Box} \bar{\varphi}(h^2-2h_{\alpha \beta}h^{\alpha \beta})]\bigg] \nonumber \\ &+ \bar{G}_{3,\varphi}\frac{\delta \varphi}{2}[\bar{\varphi}_{,\alpha}(h^{,\alpha} - 2 {h^{\alpha \beta}}_{;\beta}) + 2\bar{\Box}\delta \varphi - 2 h^{\alpha \beta} \bar{\varphi}_{;\alpha \beta} + h \bar{\Box}\bar{\varphi} ] + \bar{G}_{3,\varphi \varphi}\frac{\delta \varphi^2}{2}\bar{\Box} \bar{\varphi}\nonumber \\ & + \bar{G}_{3,XX}\frac{\bar{\Box}\bar{\varphi}}{8}\bigg[h^{\alpha \beta} \bar{ \varphi}_{,\alpha} \bar{ \varphi}_{,\beta} -2 \bar{ \varphi}_{,\gamma}\delta \varphi^{,\gamma} \bigg]^2  +\bar{G}_{3,\varphi X}\bar{\Box}\bar{\varphi}\frac{\delta \varphi}{2} [h^{\alpha \beta}\bar{\varphi}_{,\alpha}\bar{\varphi}_{,\beta} - 2 \bar{\varphi}^{,\alpha} \delta \varphi_{,\alpha}] \nonumber \\ &+\bar{G}_{3,X}\bigg[-\frac{\delta \varphi_{,\alpha}}{2}[(\delta \varphi^{,\alpha} - 2 h^{\alpha \beta}\bar{\varphi}_{,\beta})\bar{\Box}\bar{\varphi} + \bar{\varphi}^{,\alpha}(2\bar{\Box}\delta \varphi -2h^{\beta \gamma}\bar{ \varphi}_{;\beta \gamma}+h\bar{\Box}\bar{\varphi}) ] \nonumber \\ & + \frac{1}{4} \bar{\varphi}_{,\alpha}\bar{\varphi}_{,\beta}[h^{\alpha \beta}(2\bar{\Box} \delta \varphi + \bar{\varphi}_{,\gamma}(h^{,\gamma}-2{h^{\gamma \delta}}_{;\delta}) - 2 h^{\gamma \delta}\bar{\varphi}_{;\gamma \delta}+h\bar{\Box}\bar{\varphi})-2h^{\alpha \gamma}{h^{\beta}}_{ \gamma}\bar{\Box}\bar{\varphi} \nonumber \\ & \hspace{230pt}-2\delta \varphi^{,\alpha}(h^{,\beta} - 2{h^{\beta \gamma}}_{;\gamma})] \bigg]  \bigg\},
\end{align}

\begin{align}
    S^{(2)}_{4} = \frac{\sqrt{-\bar{g}}}{128\pi G}\bigg\{&\bar{G}_4\bigg[8{h_{\alpha}}^{\gamma}h^{\alpha \beta}\bar{R}_{\beta \gamma} - 4hh^{\beta \gamma}\bar{R}_{\beta \gamma} -2 h^{\alpha \beta}h_{\alpha \beta}\bar{R} + \bar{R}h^2 \nonumber \\ &  + 2(h_{,\beta}h^{,\beta} -2h^{,\beta}{h_{\beta \gamma}}^{;\gamma} -  h_{\alpha \beta;\gamma}h^{\alpha \beta;\gamma}+2h_{\alpha \gamma;\beta}h^{\alpha \beta;\gamma})\bigg]\nonumber \\ & + 4 \bar{G}_{4,\varphi}\delta \varphi[h\bar{R}-2h^{\alpha \beta}\bar{R}_{\alpha \beta} + 2({h^{\alpha \beta}}_{;\alpha \beta} - \bar{\Box}h)]+4\bar{G}_{4,\varphi \varphi}\bar{R}(\delta \varphi)^2\bigg\},
\end{align}
\end{widetext}

where $S^{(2)}_{2}$ remains as in Eq.~(\ref{S2_2}).

\section{The GW EMT} \label{sec:EMT}

\subsection{Basic variations}

We assume as independent fields of the theory $\hat{\gamma}^{\mu \nu}$, $\delta \varphi$, $\bar{\varphi}$, and $\bar{g}^{\alpha \beta}$. All other quantities can be expressed in terms of them. Before varying the action itself, we give some basic results needed.

First, we remember that (see \cite{Dinverno}, for example):
\begin{align}
\frac{\delta \bar{g}_{\mu \alpha}}{\delta \bar{g}^{\nu \lambda}} = -\frac{1}{2}( \bar{g}_{\lambda \alpha} \bar{g}_{\mu \nu} + \bar{g}_{\nu \alpha} \bar{g}_{\mu \lambda}),
\end{align}
\begin{align}
\frac{\delta \bar{g}^{\alpha \beta}}{\delta \bar{g}^{\mu \nu}} = \frac{1}{2} (\delta^{\alpha}_{\mu}\delta^{\beta}_{\nu} + \delta^{\alpha}_{\nu}\delta^{\beta}_{\mu}) \label{var_metric}
\end{align}
and
\begin{align}
    \frac{\delta \sqrt{-\bar{g}}}{\delta \bar{g}^{\alpha \beta}} = -\frac{\sqrt{-\bar{g}}}{2}\bar{g}_{\alpha \beta }. \label{var_det_g}
\end{align}
Then, one can vary all functions appearing on the Horndeski action. For example, from Eq.~(\ref{X_bar}):
\begin{align}
    \frac{\delta \bar{X}}{\delta \bar{g}^{\alpha \beta}} = -\frac{1}{2} \bar{\varphi}_{,\alpha} \bar{\varphi}_{,\beta}, \label{var_X}
\end{align}
which implies
\begin{align}
    \frac{\delta \bar{G}_i}{\delta \bar{g}^{\mu \nu}} = - \frac{\bar{G}_{i,X}}{2} \bar{ \varphi}_{,\mu}\bar{ \varphi}_{,\nu}. \label{var_G}
\end{align}

The perturbation $h^{\alpha \beta}$ is not an independent field. It is expressed in terms of the true GW degrees of freedom in Eq.~(\ref{h_in_terms_of_gamma}). Varying such relation, we find
\begin{align}
   \frac{\delta h^{\alpha \beta}}{\delta \bar{g}^{\mu \nu}} = \frac{1}{2}[\bar{g}^{\alpha \beta}\hat{\gamma}_{\mu \nu}-\hat{\gamma}\delta^{\alpha}_{(\mu}\delta^{\beta}_{\nu)}] - {\bar{A}^{\alpha \beta}}_{\mu \nu}\delta \varphi, \label{var_h}
\end{align}
where
\begin{align}
    {\bar{A}^{\alpha \beta}}_{\mu \nu} &\leqdef \frac{\delta C^{\alpha \beta}}{\delta \bar{g}^{\mu \nu}} \nonumber \\&= \frac{1}{\bar{G}_4} \bigg[-\frac{1}{2}\bar{G}_{3,XX}\bar{ \varphi}_{,\mu}\bar{ \varphi}_{,\nu}(\bar{ \varphi}^{,\alpha}\bar{ \varphi}^{,\beta} + \bar{X}\bar{g}^{\alpha \beta})\nonumber \\ & + \bar{G}_{3,X}\bigg(2\bar{ \varphi}^{(,\alpha}\delta^{\beta)}_{(\mu}\bar{ \varphi}_{,\nu)}-\frac{\bar{g}^{\alpha \beta}}{2}\bar{ \varphi}_{,\mu}\bar{ \varphi}_{,\nu} + \bar{X}\delta^{\alpha}_{(\mu} \delta^{\beta}_{\nu)}\bigg)\nonumber \\ &\hspace{150pt}+ \bar{G}_{4,\varphi}\delta^{\alpha}_{(\mu} \delta^{\beta}_{\nu)}\bigg] . \label{A}
\end{align}
Taking the trace of Eq.~(\ref{h_in_terms_of_gamma}) and varying once more:
\begin{align}
\frac{\delta h}{\delta \bar{g}^{\mu \nu}} = -\hat{\gamma}_{\mu \nu} + \bar{B}_{\mu \nu}\delta \varphi, \label{delta_h}
\end{align}
where
\begin{align}
	\bar{B}_{\mu \nu} \leqdef \frac{\delta \hat{C}}{\delta \bar{g}^{\mu \nu}} = \frac{\bar{ \varphi}_{,\mu}\bar{ \varphi}_{,\nu}}{\bar{G}_{4}}(\bar{G}_{3,X} + \bar{X}\bar{G}_{3,XX}). \label{B}
\end{align}
and $\hat{C} \leqdef \bar{g}^{\alpha \beta} \hat{C}_{\alpha \beta}$.

\subsection{Variation of covariant derivatives} \label{subsec:Variation_of covariant_derivatives}

Several terms of the action upon which we wish to operate the variation with respect to the background metric have covariant derivatives. Since these derivatives have connection coefficients, one would need, in principle, to handle the voluminous amount of variation terms arising from it. But one can show that variations of connection components vanish under the weak limit average when appearing in covariant derivatives of perturbation fields. Although this result was already obtained in \cite{Stein2011}, we revisit it to point out an important subtlety in our case. 

A typical term we are interested here and that will be present after variation of the action is:
\begin{align}
    T_1 = \epsilon^2 \int \sqrt{\bar{g}} {P^{\gamma}}_{\alpha \beta} \delta ({h^{\alpha \beta}}_{;\gamma}) d^4x, \label{term_first_derivative}
\end{align}
where $P^{\gamma}_{\alpha \beta}$ is a tensor that must be present so that the overall quantity can become a scalar present in the action. Because of the connections present in the covariant derivative, one can show that
\begin{align}
    \delta ({h^{\alpha \beta}}_{;\gamma}) = 	(\delta {h^{\alpha \beta}})_{;\gamma} + \delta \bar{\Gamma}^{\alpha}_{\delta \gamma} h^{\delta \beta} + \delta \bar{\Gamma}^{\beta}_{\delta \gamma} h^{\alpha \delta}.
\end{align}
The variation of the connection components are of the form
\begin{align}
    \delta \bar{\Gamma}^{\alpha}_{\delta \gamma} \sim \bar{g}_{\lambda \delta} (\delta \bar{g}^{\lambda \alpha})_{;\gamma}
\end{align}
So that they will appear in $T_1$ as
\begin{align}
    T_1 &\sim 	\epsilon^2 \int \sqrt{\bar{g}} P^{\gamma}_{\alpha \beta} \bar{g}_{\lambda \delta} (\delta \bar{g}^{\lambda \alpha})_{;\gamma}h^{\delta \beta} d^4x \nonumber \\&\sim - \epsilon^2 \int \sqrt{\bar{g}} (P^{\gamma}_{\alpha \beta}{h_{\lambda}}^{ \beta})_{;\gamma} \delta \bar{g}^{\lambda \alpha} d^4x.
\end{align}
which, by Eq.~(\ref{GW_EMT_definition}), results in a contribution to the GW EMT that goes like
\begin{align}
    T^{GW}_{\alpha \lambda} \sim \langle \epsilon^2({P^{\gamma}}_{\alpha \beta}{h_{\lambda}}^{\beta})_{;\gamma} \rangle.
\end{align}
By inspecting the terms in the action of the form of Eq.~(\ref{term_first_derivative}), one is able to conclude that $\mathcal{O}(\epsilon {P^{\gamma}}_{\alpha \beta}) = \mathcal{O}(\epsilon)$ or $\mathcal{O}(\epsilon^0)$. Using an analogous argument as the one given in Eq.~(\ref{derivatives_vanish}), one can conclude that the above term must vanish. This implies that we can discard connection variations when varying first covariant derivatives of the tensorial perturbation. By a similar argument, one can conclude the same property for the other perturbation derivatives, such as $\delta \phi_{;\alpha \beta}$ and $h_{\alpha \beta;\mu \nu}$.

The subtlety present in this demonstration is that it is not valid if the field being differentiated is of background nature. In theories where the metric is the only degree of freedom, this is not a problem, since covariant derivatives of the background metric are always zero. But in Horndeski theories, we have the presence of the background scalar field. Consequently, terms like
\begin{align}
    T_2 = \epsilon^2 \int \sqrt{-\bar{g}} L^{\alpha \beta} \delta(\bar{\varphi}_{;\alpha \beta}) d^4x, \label{var_background_scalar}
\end{align}
where $L^{\alpha \beta}$ is some tensor necessary to form a scalar quantity in the action, will appear after variation. Varying the correspondent connections will give contributions to the GW EMT of the form:
\begin{align}
    T_{\phi \lambda} \sim   \langle \epsilon^2 ({L^{\beta}}_{\phi}\bar{\varphi}_{,\lambda})_{;\beta} \rangle. 
\end{align}
But terms can be found in the action such that $\mathcal{O}(\epsilon^2 L^{\alpha \beta}) = \mathcal{O}(\epsilon^0)$. More explicitly, they are the ones involving:
\begin{align}
	L^{\alpha \beta} = -\frac{1}{2} \bar{G}_{3,X} (\delta \varphi)_{,\lambda} (\delta \varphi)^{,\lambda} \bar{g}^{\alpha \beta} \label{first_example}
\end{align} 
and
\begin{align}
	L^{\alpha \beta}=\frac{\bar{G}_{3,XX}}{2}(\bar{g}_{\mu \nu}\delta \varphi^{,\mu} \bar{ \varphi}^{,\nu})^2\bar{g}^{\alpha \beta}. \label{second_example}
\end{align}
In these examples, we cannot neglect the variation of connection components.

\subsection{The energy momentum tensor}

We are now finally able to obtain the GW EMT. We write the second order of the action as:
\begin{align}
	S^{(2)} = \frac{\sqrt{-\bar{g}}}{16\pi G} \sum_{i = 2}^{4} s^{(2)}_{i},
\end{align}
where $s_2^{(2)}, s_3^{(2)}$ and $s_4^{(2)}$ are related, respectively, with Eqs.~(\ref{S2_2}), (\ref{S2_3}) and (\ref{S^2_4}). We remember that, because of Eqs.~(\ref{power_drop_tensor}) and (\ref{power_drop_scalar}) and the limit made in Eq.~(\ref{weak_limit}), the only terms surviving the averaging process are those in which the number of derivatives acting on perturbation fields is at least the number of perturbation factors appearing on it. Varying and using Eq.~(\ref{GW_EMT_definition}), one can separate the GW EMT in two parts:
\begin{align}
    T^{GW}_{\mu \nu} = T^{GW(first)}_{\mu \nu} + T^{GW(second)}_{\mu \nu}.
\end{align}
The first comes from the variation of $\sqrt{-\bar{g}}$ (Eq.~(\ref{var_det_g})) and reads:
\begin{widetext}
\begin{align}
    T^{GW(first)}_{\mu \nu} &= \frac{\bar{g}_{\mu \nu}}{16\pi G}  \sum_{i = 2}^{4} \langle \epsilon^2 s^{(2)}_{i} \rangle = \frac{\bar{g}_{\mu \nu}}{16\pi G} \bigg\langle\epsilon^2\bigg\{\frac{1}{2}[\bar{G}_{2,XX}(\delta \varphi_{,\alpha} \bar{ \varphi}^{,\alpha})^2   - \bar{G}_{2,X}\delta \varphi_{,\alpha}\delta \varphi^{,\alpha}] + \bar{G}_{4,\varphi} \delta \varphi ({h^{\alpha \beta}}_{;\alpha \beta} - \bar{\Box}h)   \nonumber \\ & \hspace{130pt}+ \frac{\bar{G}_4}{4}\{h_{,\beta}h^{,\beta}  - 2 h^{,\beta}{h_{\beta \gamma}}^{;\gamma} + (2h_{\alpha \gamma;\beta}-h_{\alpha \beta;\gamma})h^{\alpha \beta;\gamma} \} + \bar{G}_{3,\varphi} \delta \varphi \bar{\Box}\delta \varphi   \nonumber \\ & \hspace{130pt}+ \bar{G}_{3,XX}\frac{\bar{\Box}\bar{ \varphi}}{2}(\bar{ \varphi}_{,\gamma}\delta \varphi^{,\gamma})^2 +\frac{1}{2}\bar{G}_{3,X}\{- \delta \varphi_{,\alpha} \delta \varphi^{,\alpha}\bar{\Box}\bar{ \varphi}   +\bar{ \varphi}_{,\alpha}\bar{ \varphi}_{,\beta}[h^{\alpha \beta}\bar{\Box}\delta \varphi \nonumber \\ & \hspace{130pt} -\delta \varphi^{,\alpha}(h^{,\beta}- 2{h^{ \beta \gamma}}_{;\gamma})] - \delta \varphi_{,\alpha} \bar{ \varphi}^{,\alpha} \bar{\Box}\delta \varphi\} \bigg\}\bigg\rangle. \label{T_first}
\end{align}
\end{widetext}
The second part comes from the variation of the $s_i^{(2)}$'s. Because variation with respect to $\bar{g}^{\alpha \beta}$ does not change the order of the quantity with respect to $\epsilon$,
\begin{align}
\bigg\langle\epsilon^2\frac{\delta s^{(2)}_i}{\delta \bar{g}^{\mu \nu}} \bigg\rangle = \frac{\delta}{\delta \bar{g}^{\mu \nu}} \langle \epsilon^2 s^{(2)}_i \rangle
\end{align}
and we only need to vary the terms that survive the averaging process.

We conclude that
\begin{align}
    T^{GW(second)}_{\mu \nu} = 
-\frac{1}{8 \pi G} \sum_{i=2}^{4} \langle \epsilon^2\delta s_i^{(2)} \rangle  = \sum_{i=2}^{4} T^{(second,i)}_{\mu \nu}, \label{T_second}
\end{align}
where we present each $T^{(second,i)}_{\mu \nu}$ and details on how to obtain them in Appendix \ref{appendix}. Combining contributions of Eq.~(\ref{T_first}) and (\ref{T_second}), we finally find:
\begin{align}
    T^{GW}_{\mu \nu} = \sum_{i=2}^{4} T^{GW(i)}_{\mu \nu}, \label{complete_EMT}
\end{align}
where
\begin{align}
    T^{GW(2)}_{\mu \nu} = \frac{1}{16\pi G}{\bar{W}^{\lambda \phi}}_{\mu \nu} \langle \epsilon^2 \delta \varphi_{,\lambda} \delta \varphi_{,\phi}\rangle, \label{T_2}
\end{align}
with
\begin{align}
    {\bar{W}^{\lambda \phi}}_{\mu \nu} =  &\bar{G}_{2,X} \bigg[\delta^{\lambda}_{\mu} \delta^{\phi}_{\nu} - \frac{\bar{g}_{\mu \nu}}{2}\bar{g}^{\lambda \phi}\bigg] \nonumber \\ & -2\bar{G}_{2,XX}\bigg[\bar{\varphi}_{(,\mu}\delta^{\phi}_{,\nu)} \bar{\varphi}^{,\lambda}+ \frac{\bar{\varphi}_{,\mu}\bar{\varphi}_{,\nu}}{4}\bar{g}^{\lambda \phi} \nonumber \\ &- \frac{\bar{g}_{\mu \nu}}{4}\bar{\varphi}^{,\lambda}\bar{\varphi}^{,\phi}\bigg] + \bar{G}_{2,XXX}\frac{\bar{\varphi}_{,\mu}\bar{\varphi}_{,\nu}}{2}\bar{\varphi}^{,\lambda}\bar{\varphi}^{,\phi},
\end{align}
\\
\begin{align}
    T^{GW(4)}_{\mu \nu} = \frac{1}{16\pi G}\bigg\{ &\frac{\bar{G}_4}{2}\bigg[\langle \epsilon^2\hat{\gamma}_{\alpha \beta;\mu}{\hat{\gamma}^{\alpha \beta}}_{;\nu}\rangle -\frac{1}{2}  \langle\epsilon^2\hat{\gamma}_{,\mu}\hat{\gamma}_{,\nu}\rangle\bigg]\nonumber \\ &+ {\bar{O}^{\lambda \phi}}_{\mu \nu} \langle \delta \varphi_{,\lambda} \delta \varphi_{,\phi}\rangle + \bar{G}_{4,\varphi}\langle \hat{\gamma}_{,\mu} \delta \varphi_{,\nu}\rangle\nonumber \\ &+  \langle N_{\mu\nu}\bar{\Box}\delta \varphi\rangle + {\bar{Q}^{\alpha \beta\lambda \phi}}_{\mu \nu} \langle \hat{\gamma}_{\alpha \beta,\lambda} \delta \varphi_{,\phi}\rangle\bigg\}, \label{T_4}
\end{align}
with
\begin{align}
    {\bar{O}^{\alpha \beta}}_{\mu \nu} = \frac{\bar{G}_4}{2}&[(C_{\gamma \lambda}C^{\gamma \lambda}-C^2)\delta^{\alpha}_{\mu} \delta^{\beta}_{\nu} - 4 {\bar{A}^{\alpha \gamma}}_{\mu \nu} {C_{\gamma}}^{\beta}\nonumber \\ &\hspace{9pt}-(C\bar{g}_{\mu \nu} + 2\bar{B}_{\mu \nu})C^{\alpha \beta}+2C{\bar{A}^{\alpha \beta}}_{\mu \nu}\nonumber \\ &\hspace{25
    pt}+(2\bar{g}_{\gamma \mu}\bar{g}_{\delta \nu}+\bar{g}_{\mu \nu}\bar{g}_{\gamma \delta})C^{\gamma\alpha}C^{\delta \beta}] \nonumber \\ & \hspace{-5pt}+ 2 \bar{G}_{4,\varphi} \bigg[C\delta_{\mu}^{\alpha}\delta_{\nu}^{\beta}+\frac{\bar{g}_{\mu \nu}}{2}C^{\alpha \beta} - {\bar{A}^{\alpha \beta}}_{\mu \nu} \bigg],
\end{align}

\begin{align}
    N_{\mu \nu} = \frac{\bar{G}_4}{2}\bigg[2h\bar{B}_{\mu \nu}+2h_{\alpha \beta} {\bar{A}^{\alpha \beta}}_{\mu \nu}&-2\bar{C}_{\mu \beta}{h_{\nu}}^{\beta} \nonumber \\ &+ \frac{\bar{g}_{\mu \nu}}{2}(hC-h_{\alpha \beta} \bar{C}^{\alpha \beta})\bigg] \nonumber \\ & \hspace{-10pt}+2\bar{G}_{4,\varphi} \delta \varphi \bigg[\bar{B}_{\mu \nu} + \frac{\bar{g}_{\mu \nu}}{2}\bar{C}\bigg], \label{N}
\end{align}
\begin{align}
    {\bar{Q}^{\alpha \beta\lambda \phi}}_{\mu \nu} = \frac{\bar{G}_4}{2}[3C^{\lambda \phi}\delta^{\alpha}_{\mu}\delta^{\beta}_{\nu}-2C^{\alpha \beta}\delta^{\lambda}_{\mu}\delta^{\phi}_{\nu}]
\end{align}
and, finally, 
\begin{align}
    T^{GW(3)}_{\mu \nu} = \hspace{-2pt}\frac{1}{16 \pi G} \hspace{-2pt}\bigg\{&{\bar{E}^{\lambda \phi}}_{\mu \nu} \langle \epsilon^2\delta \varphi_{,\lambda} \delta \varphi_{,\phi}\rangle+ \langle \epsilon^2 D_{\mu\nu}\bar{\Box}\delta \varphi\rangle\nonumber \\ & \hspace{35pt}-\frac{\bar{G}_{3,X}}{2} \bar{\varphi}_{,\gamma}\bar{\varphi}^{,\gamma} \langle \epsilon^2\hat{\gamma}_{,\mu} \delta \varphi_{,\nu}\rangle   \nonumber \\ &  \hspace{15pt} +{\bar{I}^{\alpha \beta\lambda \phi}}_{\mu \nu}  \langle \epsilon^2\hat{\gamma}_{\alpha \beta,\lambda} \delta \varphi_{,\phi}\rangle  +\bar{J}_{\mu \nu}  \nonumber \\ &\hspace{40pt}+ \bar{G}_{3,X} \bar{\varphi}^{\alpha} \langle\epsilon^2\delta \varphi_{,\alpha}\delta \varphi_{;\mu \nu}\rangle\bigg\}, \label{T_3}
\end{align}
with
\begin{align}
    D_{\mu \nu} = &\bar{G}_{3,\varphi}\bar{g}_{\mu \nu}\delta \varphi+\bar{G}_{3,\varphi X} \bar{\varphi}_{,\mu}\bar{\varphi}_{,\nu} \delta \varphi \nonumber \\  &- \bar{G}_{3,X} \bigg[\bar{\varphi}_{,\alpha}\bar{\varphi}_{,\beta} \frac{\delta h^{\alpha \beta}}{\delta \bar{g}^{\mu \nu}} + \bar{\varphi}_{;\mu \nu}\delta \varphi - \bar{\varphi}_{(,\mu}\delta \varphi_{,\nu)}\nonumber \\ &\hspace{45pt}-\frac{\bar{g}_{\mu \nu}}{2}(\delta \varphi \bar{\Box}\bar{\varphi}+\bar{\varphi}_{,\alpha}\bar{\varphi}_{,\beta} h^{\alpha \beta}-\delta \varphi_{,\alpha}\bar{\varphi}^{,\alpha})\bigg] \nonumber \\ &- \bar{G}_{3,XX} \bigg[\frac{\bar{\varphi}_{,\mu}\bar{\varphi}_{,\nu}}{2}(\delta \varphi_{,\alpha}\bar{\varphi}^{,\alpha} - \delta \varphi \bar{\Box}\bar{\varphi} - \bar{\varphi}_{,\alpha}\bar{\varphi}_{,\beta}h^{\alpha \beta})\bigg]\nonumber \\ & \hspace{90pt}+ \bar{G}_{3,XXX} \frac{\bar{\varphi}_{,\mu}\bar{\varphi}_{,\nu}}{2}(\bar{\varphi}_{,\gamma}\delta \varphi^{,\gamma})^2,
\end{align}
\begin{align}
    {\bar{E}^{\lambda \phi}}_{\mu \nu} = &2 \bar{G}_{3,\varphi} \delta^{\lambda}_{\mu}\delta^{\phi}_{\nu} \nonumber \\ & +\bar{G}_{3,X} \bigg\{[ \bar{\Box}\bar{\varphi}-\bar{C}^{\alpha \beta}\bar{\varphi}_{\alpha}\bar{\varphi}_{\beta}]\delta^{\lambda}_{\mu}\delta^{\phi}_{\nu} \nonumber \\ &  +2 \bar{\varphi}_{,\beta} \bar{\varphi}_{(,\mu} \delta^{\lambda}_{\nu)} \bar{C}^{\beta \phi}  +\frac{\bar{g}_{\mu \nu}}{2}\bar{\varphi}^{,\lambda}[\bar{C}\bar{\varphi}^{,\phi} - 2 \bar{C}^{\beta \phi}\bar{\varphi}_{,\beta}] \nonumber \\ & \hspace{20pt} -\bar{\varphi}^{,\lambda} [2\bar{C}\delta^{\phi}_{(\mu}\bar{\varphi}_{,\nu)} - \bar{B}_{\mu \nu} \bar{\varphi}^{,\phi} - 2 {\bar{A}^{\beta \phi}}_{\mu \nu}\bar{\varphi}_{,\beta}] \bigg\} \nonumber \\ &-2 \bar{G}_{3,XX} \bar{\varphi}^{,\lambda}\bigg\{\bar{\varphi}_{(,\mu}\delta^{\phi}_{\nu)}\bar{\Box} \bar{\varphi} -\frac{1}{2}\bar{\varphi}^{,\phi}\bar{\varphi}_{;\mu\nu} \nonumber \\ & \hspace{20pt}- \frac{\bar{g}_{\mu \nu}}{4}\bar{\varphi}^{\phi} \bar{\Box}\bar{\varphi}  - \frac{\bar{\varphi}_{,\mu}\bar{\varphi}_{,\nu}}{4}[C\bar{\varphi}^{,\phi} - 2 \bar{C}^{\beta \phi}\bar{\varphi}_{,\beta}] \bigg\},
\end{align}
\begin{align}
    {\bar{I}^{\alpha \beta \lambda \phi}}_{\mu \nu} = &\bar{G}_{3,X} (\bar{\varphi}^{,\alpha}\bar{\varphi}^{,\beta}\delta^{\lambda}_{\mu}\delta^{\phi}_{\nu} - 2 \bar{\varphi}^{,\lambda}\bar{\varphi}^{,\phi}\delta^{\alpha}_{\mu}\delta^{\beta}_{\nu}),
\end{align}
\begin{align}
    \bar{J}_{\mu \nu} = & \bigg(\delta^{\alpha}_{(\mu}\delta^{\beta}_{\nu)}-\bar{g}^{\alpha \beta}\frac{\bar{g}_{\mu \nu}}{2}\bigg)\langle [(\bar{G}_{3,XX}(\bar{\varphi}_{,\gamma}\delta \varphi^{,\gamma})^2\nonumber \\ &\hspace{86pt}-\bar{G}_{3,X} \delta \varphi_{,\gamma} \delta \varphi^{,\gamma})\bar{\varphi}_{,\alpha}]_{;\beta}\rangle 
\end{align}

Because the quantities $\bar{C}_{\mu \nu}$, $\bar{B}_{\mu \nu}$ and ${\bar{A}^{\alpha \beta}}_{\mu \nu}$ are totally symmetric in their indices and because of the property exemplified by Eq.~(\ref{derivative_exchange}), almost all terms appearing in the GW EMT are symmetric at a glance. The only term not manifestly symmetric is $-\langle \bar{G}_4 \bar{C}_{\mu \beta} {h_{\nu}}^{\beta} \bar{\Box} \delta \varphi \rangle$, present in Eq.~(\ref{N}). It is simple to notice, using the properties derived in Subsec.~(\ref{subsec:weak_limit}), that:
\begin{align}
    \langle \bar{G}_4 \bar{C}_{\mu \beta} {h_{\nu}}^{\beta} \bar{\Box} \delta \varphi \rangle = \langle \bar{G}_4  {h_{\nu}}^{\beta} \bar{\Box} (\bar{C}_{\mu \beta} \delta \varphi) \rangle 
\end{align}
since the difference between the sides of the equation vanishes in the weak limit. Furthermore, since we are assuming luminal tensorial modes, Eq.~(\ref{dalambertian_vanish_tensor}), one may add a vanishing quantity to obtain:
\begin{align}
     \langle \bar{G}_4 \bar{C}_{\mu \beta} {h_{\nu}}^{\beta} \bar{\Box} \delta \varphi \rangle &= - \langle \bar{G}_4  {h_{\nu}}^{\beta} \bar{\Box} (\hat{\gamma}_{\mu \beta}-\bar{C}_{\mu \beta} \delta \varphi) \rangle \nonumber \\ & = - \bigg\langle \bar{G}_4  {h_{\nu}}^{\beta} \bar{\Box} h_{\mu \beta}+\frac{1}{2}h_{\mu\nu}\bar{\Box}\hat{\gamma} \bigg\rangle,
\end{align}
where in the last equality we have used Eq.~(\ref{h_in_terms_of_gamma}).
While the second term is evidently symmetric, the first one is shown to be as well by noticing that:
\begin{align}
    \langle \bar{G}_4 {h_{\nu}}^{\beta} \bar{\Box} h_{\mu \beta} \rangle = \langle \bar{G}_4  h_{\mu \beta} \bar{\Box}{h_{\nu}}^{\beta}  \rangle,
\end{align}
a result of the property exemplified by Eq.~(\ref{by_parts}). We then can conclude that 
\begin{align}
    T^{(GW)}_{\mu \nu} = T^{(GW)}_{\nu \mu}, 
\end{align}
a property expected either from physical reasons or simply from the definition of GW EMT, Eq.~(\ref{EMT_def}).

Another important aspect we would like to investigate is the trace of the tensor. In GR, GWs behave like a radiation fluid, since the trace of its EMT vanishes. But in more general theories of gravity, the tensor can have non-vanishing trace. In particular, if such quantity is negative, as studies in $f(R)$ theories already have suggested to be the case \cite{Preston2016}, it can effectively act as a dark energy candidate in the context of cosmological space-times.   

We can inquire if the trace of the obtained GW EMT vanishes in general or if there is the possibility of the correspondent fluid to have distinct behaviors depending on the subclass of theories one is interested in. If one wants to prove that such trace does not vanish in the general reduced Horndeski scheme, it is only necessary to show that the trace of a particular coefficient of the $\bar{G}_i$ 
functions does not vanish, since they are independent from each other. 

Assuming, for example, that $\bar{G}_{2,X} \neq 0$ but $\bar{G}_{2,XX} =0$:
\begin{align}
    T^{GW(2)} \leqdef {{T^{GW(2)}}_{\mu}}^{\mu} &= -\frac{1}{16\pi G} \langle \epsilon^2\delta \varphi _{,\mu} \delta \varphi^{,\mu} \rangle \nonumber \\ &=  \frac{1}{16\pi G} \langle \epsilon^2\delta \varphi \bar{\Box} \delta \varphi \rangle,
\end{align}
which is not necessarily zero for theories where scalar waves are non-luminal. We then conclude that there are theories within the reduce Horndeski family that have the GW EMT with trace different from zero. It would be interesting to investigate, for example, in what subclass of theories this tensor has negative trace and, thus, can behave as dark energy.

We additionally point out that, although the weak limit average discards all contributions of order $\epsilon^n$, with $n>0$, it amplifies terms having negative powers in this perturbation factor. They are terms in which the number of derivatives acting on perturbations exceeds the number of perturbation factors. In the tensor calculated, there are terms of this kind only in $T^{GW(3)}_{\mu \nu}$, more precisely, in $D_{\mu \nu}$ and in the last term of Eq.~(\ref{T_3}). The $\bar{J}_{\mu \nu}$ contribution, although having more derivatives than perturbations, is not problematic since one of the derivatives is a total derivative and, by the property established in Eq.~{\ref{average_derivative_vanish}}, it has order $\epsilon^0$. The terms with negative powers in $\epsilon$ have divergent behavior under the weak limit and, if present, make the GW EMT loose its physical meaning. 

We emphasize that the complete equations of motion were not imposed on the GW EMT and that they may be helpful when dealing with such divergent terms. For example, it is easy to see that all divergent terms vanish when we deal with luminal scalar-waves, since the dominant order in powers of $\epsilon$ of the equation of motion make D'alembertian terms vanish in the EMT and $\bar{G}_{3,X} = 0$. Furthermore, theories in which $\bar{G}_{3,X} = 0$ (a necessary but not sufficient condition for luminal scalar-waves) do not have any divergent term. On the other hand, we believe that, in theories in which such terms do not vanish after imposing the equations of motion, the procedure adopted here to obtain the energy-momentum information of GWs is incomplete and another formalism must be developed in order to treat it. A possible solution to this stalemate is to maintain the amplitude and wavelength perturbative parameters ($\alpha$ and $\epsilon$) initially introduced in Eq.~(\ref{background_wave_split}) as independent variables (instead of assuming them to be of same order, as in Eq.(\ref{amplitude_wavelength})) and to make a limit procedure in each one, imposing a hierarchy of smallness between them.

\section{Special cases} \label{sec:special cases}

\begin{figure}[!h]
	\centering
	\includegraphics[width=9.5
	cm]{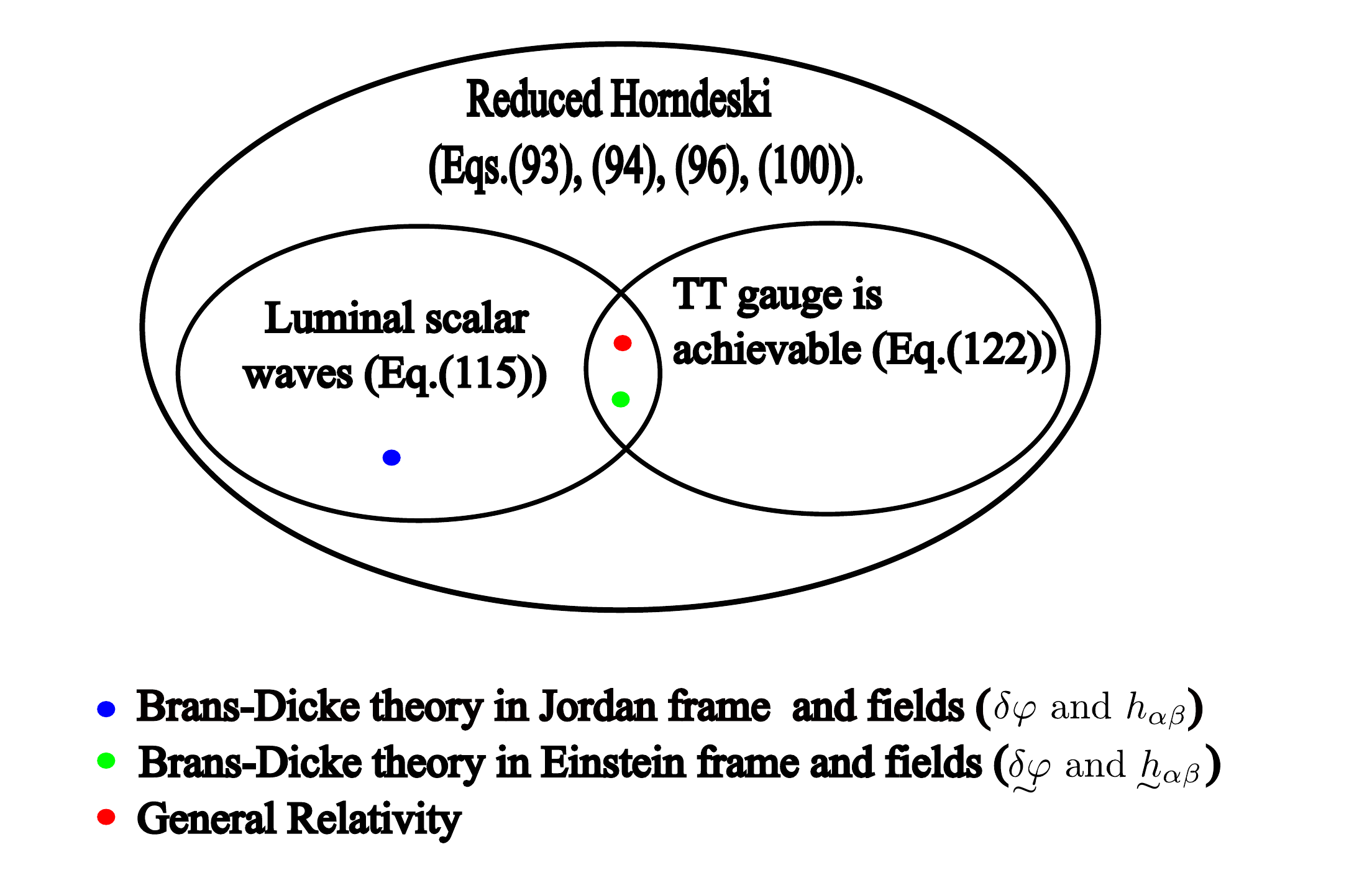}
	\caption{Special cases studied in this work, with the equations that give the corresponding GW EMT. Dots stand for particular theories and where they fit in the diagram.}
	\label{summary}
\end{figure}

We now investigate particular important cases of the GW EMT. We take the opportunity to reobtain results already derived in literature regarding the particular case of a Brans-Dicke theory. The special cases are summarized in Fig.(\ref{summary}).

\subsection{Luminal scalar waves}

Scalar waves become luminal when the diagonalized version of the kinetic operator present in Eq.~(\ref{coupled_GW_equations}) is proportional to the metric which, consequently, implies that the second order differential operator acting on the scalar perturbation is just the D'alembertian. As pointed out in \cite{Dalang2021}, this occurs in theories where:
\begin{align}
    \bar{G}_{2,XX} = \bar{G}_{3,X} = 0.
\end{align}
Inserting this equation in Eqs.~(\ref{C}), (\ref{A}) and (\ref{B}), we find:
\begin{align}
    {\bar{A}^{\alpha \beta}}_{\mu \nu} &= \frac{\bar{G}_{4,\varphi}}{\bar{G}_4} \delta^{\alpha}_{(\mu} \delta^{\beta}_{\nu)}, \\ 
    \bar{B}_{\mu \nu} &= 0, \\
    \bar{C}_{\mu \nu} &= \frac{\bar{G}_{4,\varphi}}{\bar{G}_4} \bar{g}_{\mu \nu}.
\end{align}

Analogous to Eq.~(\ref{dalambertian_vanish_tensor}), the dominant term of the equation of motion in orders of $\epsilon$ for scalar waves imply, then:
\begin{align}
    \langle \epsilon^2 W \bar{\Box} \delta \varphi \rangle = 0,
\end{align}
where $W = \mathcal{O}(1)$.

With these simplifications, the GW EMT becomes (remembering the integration by parts property exemplified in Eq.~(\ref{by_parts})):
\begin{align}
    T^{GW}_{\mu \nu} = \frac{1}{16\pi G} \bigg\{&\bigg(\bar{G}_{2,X}+2\bar{G}_{3,\varphi}+3 \frac{\bar{G}_{4,\varphi}^2}{\bar{G}_4}\bigg) \langle\epsilon^2\delta \varphi_{,\mu} \delta \varphi_{,\nu}\rangle \nonumber \\ &+\frac{\bar{G}_4}{2}\bigg[\langle \epsilon^2\hat{\gamma}_{\alpha \beta;\mu}{\hat{\gamma}^{\alpha \beta}}_{;\nu}\rangle -\frac{1}{2}  \langle\epsilon^2\hat{\gamma}_{,\mu}\hat{\gamma}_{,\nu}\rangle\bigg] \bigg\}. \label{luminal_GW_EMT}
\end{align}

Since in this subclass the D'alembertian of every perturbation vanishes, the trace of the GW EMT continue to be zero:
\begin{align}
    T^{GW} \leqdef T^{GW \mu}_{\mu} = 0. \label{vanishing_trace}
\end{align}
which is expected since here all fields travel at the speed of light and the correspondent fluid should be of radiation nature as a consequence.

\subsection{Theories with TT gauge}

As mentioned previously, although the harmonic gauge can always be achieved, this is not true to the more restrict TT gauge. By the arguments given in \cite{Dalang2021}, the TT gauge can only be obtained in an open set if \footnote{The condition here should be $G_{3,X}=G_{4,\varphi}=0$. Expanding these in power of $\epsilon$ and demanding that the $\epsilon^0$ contributions must vanish alone, gives the correspondent statements for the background quantities.}:
\begin{align}
    \bar{G}_{3,X} = \bar{G}_{4,\varphi} = 0.
\end{align}

In this case:
\begin{align}
    &\bar{C}_{\mu \nu} = \bar{B}_{\mu \nu} = 0, \\
    &{\bar{A}^{\alpha \beta}}_{\mu \nu} = 0.
\end{align}
Since $C_{\mu \nu} = 0$, 
\begin{align}
    \hat{\gamma}_{\mu \nu} = \hat{h}_{\mu \nu}. \label{TT_gauge_gamma}
\end{align}
Since the TT gauge is achievable, we demand that:
\begin{align}
    \hat{h} \leqdef {\hat{h}^{\alpha}}_{\alpha} = 0.
\end{align}

The GW EMT gives:

\begin{align}
    T^{GW}_{\mu \nu} = T^{GW (2)}_{\mu \nu} + &\frac{1}{16 \pi G}\bigg\{ \frac{\bar{G}_4}{2}\langle \epsilon^2{\hat{\gamma}_{\alpha \beta}}_{;\mu} {\hat{\gamma}^{\alpha \beta}}_{;\nu} \rangle \nonumber \\ &+ \bar{G}_{3,\varphi}[2\delta^{\lambda}_{\mu}\delta^{\phi}_{\nu} - \bar{g}_{\mu \nu}\bar{g}^{\lambda \phi}]\langle \epsilon^2\delta \varphi_{,\lambda} \delta \varphi_{,\phi}\rangle\bigg\}. \label{TT_EMT}
\end{align}
In this case, the trace is not trivially zero:
\begin{align}
    T^{GW} = &[-\bar{G}_{2,X} \bar{g}^{\lambda \phi}- \bar{X}(\bar{G}_{2,XXX}\bar{\varphi}^{,\lambda}\bar{\varphi}^{,\phi}-\bar{G}_{2,XX}\bar{g}^{\lambda \phi}) \nonumber \\ &\hspace{97pt}-2\bar{G}_{3,\varphi}\bar{g}^{\lambda \phi}]\langle  \epsilon^2\delta \varphi_{,\lambda} \delta \varphi_{,\phi}\rangle. \label{TT_gauge}
\end{align}

\subsection{Brans-Dicke theory}

For the Brans-Dicke theory, the GW EMT and the consequent signal for the stochastic gravitational-wave background were already obtained in \cite{Isi2018}, for the Einstein frame. Here we reobtain it as a particular case of our result and obtain the correspondent Jordan frame tensor as well, comparing both. 

The action of Brans-Dicke theory in the Jordan frame is:
\begin{align}
	S^{(BD)} = \frac{1}{16 \pi G} \int \sqrt{-g} \bigg[\varphi R - \frac{\omega_{BD}}{\varphi}g^{\alpha \beta}\varphi_{,\alpha} \varphi_{,\beta}\bigg] d^4x,
\end{align}
where $\omega_{BD}$ is a real number.
This is a special case of reduced Horndeski theory with
\begin{align}
G_2(\varphi, X) &= 2\frac{\omega_{BD}}{\varphi} X, \label{G2_Jordan}\\
	G_4(\varphi, X) &= \varphi, \label{G4_Jordan}\\
	G_3(\varphi, X) &= G_5(\varphi, X) = 0\label{G3_Jordan}.
\end{align}

It is then a theory in which scalar waves are luminal ($G_{3,X} = G_{2,XX}=0$) but the TT gauge cannot be achieved ($G_{4,\varphi} \neq 0$).
From Eq.~(\ref{luminal_GW_EMT}), then, the GW EMT in the Jordan frame is:
\begin{align}
    	16 \pi G T_{\mu \nu}^{(BD} = \frac{(2\omega_{BD}+3)}{\bar{\varphi}} \langle \epsilon^2\delta \varphi_{,\mu} \delta \varphi_{,\nu} \rangle  &-\frac{\bar{\varphi}}{4} \langle \epsilon^2\hat{\gamma}_{;\mu}\hat{\gamma}_{;\nu}\rangle \nonumber \\ &\hspace{-10pt}+ \frac{\bar{\varphi}}{2} \langle \epsilon^2\hat{\gamma}_{\alpha \beta;\mu}{\hat{\gamma}^{\alpha \beta}}_{;\nu}\rangle . \label{Jordan_EMT}
\end{align}

One usually states the theory in terms of another action, that is more similar to the GR one: 
\begin{align}
\undertilde{S}^{(BD)} = \frac{1}{16 \pi G} \int \sqrt{-\undertilde{g}} [\undertilde{R}-2\undertilde{g}^{\alpha \beta}\undertilde{\varphi}_{,\alpha} \undertilde{\varphi}_{,\beta}] d^4x,
\end{align}
where the new metric and the new scalar field are defined as:
\begin{align}	&\undertilde{g}_{\mu \nu} \leqdef \frac{\varphi}{\bar{\varphi}}g_{\mu \nu}, \label{gtilde}\\
&\undertilde{\varphi} \leqdef \undertilde{\bar{\varphi}} + \epsilon \undertilde{\delta \varphi}.
\end{align}
with
\begin{align}
    \frac{\varphi}{\bar{\varphi}} &\reqdef e^{-2\frac{\epsilon\undertilde{\delta \varphi}}{\sqrt{3+2\omega_{BD}}}}. \label{phi}
\end{align}
When such transformation is made, we say we have changed from the Jordan to the Einstein frame. Undertilde quantities indicate that the Einstein frame fields are being used.

Expanding both sides of Eq.~(\ref{gtilde}), one finds:
\begin{align}
\undertilde{\bar{g}}_{\mu \nu} &= \bar{g}_{\mu \nu} , \label{same_background_frames}\\
\undertilde{h}_{\mu \nu} &= h_{\mu \nu} + \frac{\delta \varphi}{\bar{\varphi}}\bar{g}_{\mu \nu}. \label{h_frame_change}
\end{align}
Because of Eq.~(\ref{same_background_frames}), one does not need to distinguish between contractions or covariant derivatives with one or the other background metric.
Expanding Eq.~(\ref{phi}) to linear order in $\epsilon$:
\begin{align}
	\undertilde{\delta \varphi} = - \frac{\delta \varphi}{\bar{\varphi}}\frac{\sqrt{3 + 2\omega_{BD}}}{2}. \label{scalar_frame_transformation}
\end{align}

Notice that, assuming the Einstein frame fields to be the fundamental ones, the Horndeski functions would be:
\begin{align}
\undertilde{G}_2(\undertilde{\varphi}, \undertilde{X}) &= 4 \undertilde{X},\\
	\undertilde{G}_4(\undertilde{\varphi}, \undertilde{X}) &= 1,\\
	\undertilde{G}_3(\undertilde{\varphi}, \undertilde{X}) &= \undertilde{G}_5(\undertilde{\varphi}, \undertilde{X}) = 0.
\end{align}
We see that in this case, scalar waves continue to be luminal, but now the TT gauge can be achieved, since $\undertilde{G}_{4,\undertilde{\phi}} = 0 = \undertilde{G}_{3,\undertilde{X}}$.

The Einstein frame GW EMT is, then:
\begin{align}
	16 \pi G\undertilde{T}^{(BD)}_{\mu \nu} = 4\langle \epsilon^2\undertilde{\delta  \varphi}_{,\mu} \undertilde{\delta \varphi}_{,\nu}\rangle &+ \frac{1}{2} \langle   \epsilon^2\undertilde{\hat{h}}_{\alpha \beta;\mu}{\undertilde{\hat{h}}^{\alpha \beta}}_{;\nu} \rangle \nonumber \\ &- \frac{1}{4} \langle\epsilon^2\undertilde{\hat{h}}_{,\mu}\undertilde{\hat{h}}_{,\nu}\rangle, \label{EMT_Einstein_frame}
\end{align}
where we have used that Eq.~(\ref{TT_gauge_gamma}) is valid on the Einstein frame but have not imposed the TT gauge \footnote{Notice that no TT gauge is necessary to obtain Eq.~(\ref{TT_gauge_gamma}). It is only necessary for the theory to have the right Horndeski functions so that the TT gauge is achievable. The gauge we are using in this section is the given by Eq.~(\ref{gauge_bd}).} so that the comparison with literature can be done. The next step is to use Eqs.~(\ref{h_frame_change}) and (\ref{scalar_frame_transformation}) to express everything in terms of the Jordan fields. We have:
\begin{align}
	4\langle \delta  \undertilde{\varphi}_{,\mu} \delta \undertilde{\varphi}_{,\nu}\rangle = \frac{3 + 2\omega_{BD}}{\bar{\varphi}^2}\langle \delta  \varphi_{,\mu} \delta \varphi_{,\nu}\rangle \label{Einstein_to_Jordan_scalar_covariance}
\end{align}
and
\begin{align}
\undertilde{\hat{h}}^{\alpha \beta} &\leqdef \undertilde{h}^{\alpha \beta} - \bar{g}^{\alpha \beta} \frac{\undertilde{h}}{2} \nonumber \\ &= h^{\alpha \beta} - \frac{\delta \varphi}{\bar{\varphi}} \bar{g}^{\alpha \beta} - \bar{g}^{\alpha \beta} \frac{h}{2}. \label{Einstein_to_Jordan_tensor_covariance}
\end{align}

The GW EMT for Brans-Dicke was obtained in \cite{Isi2018}. The gauge used there is such that:
\begin{align}
	h = - 2 \frac{\delta \varphi}{ \bar{\varphi}}. \label{gauge_bd}
\end{align}
Which implies, then, by Eq.~(\ref{Einstein_to_Jordan_tensor_covariance}),
\begin{align}
	\undertilde{\hat{h}}^{\alpha \beta} = h^{\alpha \beta}, \label{hs_in_two_frames}
\end{align}
so that the Einstein frame EMT is expressed by the Jordan frame fields as:
\begin{align}
16 \pi G\undertilde{T}^{(BD)}_{\mu \nu} = \frac{3 + 2\omega_{BD}}{\bar{\varphi}^2}\langle \epsilon^2\delta  \varphi_{,\mu} \delta \varphi_{,\nu}\rangle &+ \frac{1}{2} \langle   \epsilon^2h_{\alpha \beta;\mu}{h^{\alpha \beta}}_{;\nu}\rangle \nonumber \\ &\hspace{20pt}- \frac{1}{4} \langle \epsilon^2 h_{,\mu} h_{,\nu} \rangle.
\end{align}
Finally, using Eq.~(\ref{gauge_bd}) again,
\begin{align}
	16 \pi G\undertilde{T}^{(BD)}_{\mu \nu} = \frac{2(1 + \omega_{BD})}{\bar{\varphi}^2}\langle \epsilon^2\delta  \varphi_{,\mu} \delta \varphi_{,\nu}\rangle \nonumber \\ \hspace{120pt}+ \frac{1}{2} \langle   \epsilon^2 h_{\alpha \beta;\mu}{h^{\alpha \beta}}_{;\nu}\rangle. \label{Einstein_EMT}
\end{align}
Which coincides with the result obtained in \cite{Isi2018}.

Since we have computed the GW EMT in both frames as special cases of our result, we can compare them. To do so, one must express the Einstein EMT in terms of $\hat{\gamma}_{\alpha \beta}$. With Eqs.~(\ref{G2_Jordan}-\ref{G3_Jordan}), Eq.~(\ref{h_in_terms_of_gamma}) becomes: 
\begin{align}
    h_{\alpha \beta} = \hat{\gamma}_{\alpha \beta} - \frac{\bar{g}_{\alpha \beta}}{2} \hat{\gamma} - \frac{\bar{g}_{\alpha \beta}}{\bar{\varphi}} \delta \varphi
\end{align}
Using the gauge condition of Eq.~(\ref{gauge_bd}) in the trace of the above equation:
\begin{align}
    \hat{\gamma} = -2 \frac{\delta \varphi}{\bar{\varphi}}, \label{gauge_bd_2}
\end{align}
which implies in
\begin{align}
    h_{\alpha \beta} = \hat{\gamma}_{\alpha \beta}. \label{gamma_equal_h}
\end{align}
Since Eq.~(\ref{TT_gauge_gamma}) is valid in the Einstein frame and Eq.~(\ref{hs_in_two_frames}) is also valid, we also conclude that
\begin{align}\undertilde{\hat\gamma}_{\alpha \beta}=\hat{\gamma}_{\alpha \beta},
\end{align}
that is, although the $h_{\alpha \beta}$ in both frames are not equal to each other, the true tensorial degrees of freedom $\hat{\gamma}_{\alpha \beta}$ are.

Using the gauge of Eq.~(\ref{gauge_bd_2}) in the Jordan frame GW EMT given by Eq.~(\ref{Jordan_EMT}) and comparing the result with the Einstein frame EMT given by Eq.~{\ref{Einstein_EMT}) with Eq.~(\ref{gamma_equal_h}) replaced on it:
\begin{align}
    T^{GW(BD)}_{\mu \nu} = \bar{\varphi} \undertilde{T}^{GW(BD)}_{\mu \nu}. \label{Jordan_to_Einstein_EMT}
\end{align}

This is not the same relation  the matter EMT $T_{\mu \nu}^{(m)}$ would have under conformal transformation \cite{Capozziello2011}:
\begin{align}
    T_{ \mu \nu}^{(m)} = \frac{\varphi}{\bar{\varphi}}\undertilde{T}_{\mu \nu}^{(m)}
\end{align}
Requiring further that four-velocities must have norm equal to $-1$, they relate as:
\begin{align}
    u^{\alpha} = \sqrt{\frac{\varphi}{\bar{\varphi}}}\undertilde{u}^{\alpha}.
\end{align}
The energy density of GWs in both frames relate, then, as:
\begin{align}
    \rho^{GW}\leqdef u^{\alpha} u^{\beta} T^{GW}_{\alpha \beta} = \varphi \undertilde{u}^{\alpha} \undertilde{u}^{\beta} \undertilde{T}^{GW}_{\alpha \beta} = \varphi \undertilde{\rho}^{GW} \label{Jordan_to_Einstein_energy}
\end{align}
which implies that, in principle, different from what is sustained in \cite{Isi2018}, the GW energy density is not conformally invariant. 
\section{Divergence of the GW EMT} \label{sec: divergence}

As we have demonstrated, some features of the GW EMT that are valid in GR, are not valid in the general context of reduced Horndeski theories, such as the trace being zero. We know that in GR, the divergence of this tensor vanishes, which implies in conservation of the number of gravitons. We want now to investigate if this is still true in our scope. 

In this section, we will work with solutions of the eikonal form. That is:
\begin{align}
&\hat{\gamma}_{\alpha \beta} = \Gamma_{\alpha \beta}(x^{\mu}) e^{iw(x^{\mu})/\epsilon}, \label{tensor_eikonal}\\
& \delta \varphi = \Phi(x^{\mu}) e^{iv(x^{\mu})/\epsilon}, \label{scalar_eikonal}
\end{align}
where $\Gamma_{\alpha \beta}, \Phi$ are complex functions and $w,v$ are real.

We choose a particular theory given by conditions
\begin{align}
    \bar{G}_2 = \bar{G}_{3,X} = \bar{G}_{4,\varphi} = \bar{G}_{4,X} = 0. \label{particular_theory}
\end{align}
This theory has luminal scalar waves and admits the TT gauge. Its GW EMT reads:

\begin{align}
    T^{GW}_{\mu \nu} =  &\frac{1}{16 \pi G}\bigg\{ \frac{\bar{G}_4}{2}\langle \epsilon^2{\hat{\gamma}_{\alpha \beta}}_{;\mu} ({\hat{\gamma}^{\alpha \beta}}_{;\nu})^* \rangle \nonumber \\ &\hspace{50pt}+ 2\bar{G}_{3,\varphi}\langle \epsilon^2 \delta \varphi_{,\mu} (\delta \varphi_{,\nu})^*\rangle\bigg\}. \label{particular_EMT}
\end{align}
where the star stands for complex conjugate. It is present on the above equation so that the final tensor remains real.

Taking the divergence of first term and replacing Eq.(\ref{tensor_eikonal}), one finds:
\begin{align}
    \bigg[\frac{\bar{G}_4}{2} \langle \epsilon^2\hat{\gamma}_{\alpha \beta;\mu} ({\hat{\gamma}^{\alpha \beta}}_{;\nu})^*\rangle\bigg]^{;\nu} = \frac{\bar{G}_{4}}{2} (k_{\mu} k_{\nu} \Gamma^2 )^{;\nu}, \label{tensor_divergence_in_eikonal} 
\end{align}
where all other contributions vanish under the weak limit average and 
\begin{align}
    \Gamma^2 \leqdef \Gamma_{\alpha \beta} (\Gamma^{\alpha \beta})^*.
\end{align}
and
\begin{align}
    k_{\mu} \leqdef w_{,\mu}.
\end{align}

We now use the equations of motion. The $\epsilon^{-1}$ contribution of the diagonalized version of Eq.~(\ref{coupled_GW_equations}) gives, for the tensorial part, in any reduced Horndeski theory \cite{Dalang2021}:
\begin{align}
k_{\mu}k^{\mu} = 0.
\end{align}
Meaning that the tensorial waves travel through null curves. Using the definition of $k_{\mu}$, the above equation implies that this curve is also a geodesic, namely:
\begin{align}
    k^{\nu} k_{\mu;\nu} = 0. \label{geodesic}
\end{align}
Furthermore, the terms of order 1 in $\epsilon$ of the equation of motion combine to give, in the particular theory where Eq.(\ref{particular_theory}) is valid, \cite{Dalang2021} 
\begin{align}
\frac{\bar{G}_4}{2}(2k^{\alpha}\Gamma_{\mu \nu;\alpha} + {k^{\alpha}}_{;\alpha}\Gamma_{\mu \nu}) = \bar{G}_{3,\varphi}\Phi e^{i(v-w)}(&\bar{\varphi}^{,\gamma}q_{\gamma} \bar{g}_{\mu \nu} \nonumber \\ &- 2\bar{\varphi}_{(,\mu} q_{\nu)}).
\end{align}
where
\begin{align}
    q_{\mu} \leqdef v_{,\mu}.
\end{align}
Contracting the equation with $(\Gamma^{\mu \nu})^*$ and summing with its complex conjugate contracted with $\Gamma^{\mu \nu}$, one is able to write
\begin{align}
    \frac{\bar{G}_4}{2}(k^{\alpha} \Gamma^2)_{; \alpha} = -2\bar{G}_{3,\varphi}\bar{\varphi}_{,\mu} q_{\nu} \text{Re}(\Phi e^{i(v-w)}  (\Gamma^{\mu \nu})^*)\label{square_amplitude_evolution}
\end{align}
where Re(...) indicates the real part of a function and we have used that, under the TT gauge we are assuming, $\Gamma_{\mu \nu}$ has vanishing trace.

Using Eqs.~(\ref{geodesic}) and (\ref{square_amplitude_evolution})
on Eq.~(\ref{tensor_divergence_in_eikonal}), one ends up with:
\begin{align}
\bigg[\frac{\bar{G}_4}{2} \langle \epsilon^2\hat{\gamma}_{\alpha \beta;\mu} ({\hat{\gamma}^{\alpha \beta}}_{;\nu})^*\rangle\bigg]^{;\nu} = &-2  \bar{G}_{3,\varphi} \bar{\varphi}_{,\alpha} q_{\beta}k_{\mu}\times \nonumber \\ &\hspace{25pt}\times\text{Re}(\Phi (\Gamma^{\alpha \beta})^*e^{i(v-w)}). \label{first_term_divergence}
\end{align}
Since $\bar{G}_{3,\varphi} = 0$ in GR, we would conclude that the GW EMT divergence vanishes in this case. But here this part doesn't vanish, even under the TT gauge and after the imposition of the equations of motion.

We proceed on computing the divergence of the second term in Eq.~(\ref{particular_EMT}). Replacing Eq.~(\ref{scalar_eikonal}) on it:
\begin{align}
    [2\bar{G}_{3,\varphi} \langle \epsilon^2 \delta \varphi_{,\mu} (\delta \varphi_{,\nu})^*\rangle]^{;\nu} = 2(\bar{G}_{3,\varphi}q_{\mu}q_{\nu}\Phi^2)^{;\nu}
\end{align}
where 
\begin{align}
    \Phi^2 \leqdef \Phi \Phi^*.
\end{align}

As in the case for tensorial waves, here we have, once more, that the dominant term of the field equation for the perturbations give \cite{Dalang2021}:
\begin{align}
    q_{\alpha}q^{\alpha} =0,
\end{align}
which implies in 
\begin{align}
    q^{\alpha}{q_{\mu}}_{;\alpha} = 0. \label{scalar_geodesic}
\end{align}
Furthermore, the amplitude evolution for the scalar waves become \cite{Dalang2021}:
\begin{align}
    2 \bar{G}_{3,\varphi} (2q^{\alpha}\Phi_{,\alpha}+{q^{\alpha}}_{;\alpha}\Phi) + 2\bar{G}_{3,\varphi \varphi}\bar{\varphi}^{,\alpha}q_{\alpha}\Phi=0 
\end{align}
Multiplying the equation by $\Phi^*$ and summing with its complex conjugate multiplied by $\Phi$, one obtains:
\begin{align}
(\bar{G}_{3,\varphi}q^{\alpha}\Phi^2)_{;\alpha} = 0.\label{scalar_amplitude_evolution}
\end{align}

Using Eqs.~(\ref{scalar_geodesic}) and (\ref{scalar_amplitude_evolution}), one concludes that:
\begin{align}
     [2\bar{G}_{3,\varphi} \langle \delta \varphi_{,\mu} (\delta \varphi_{,\nu})^*\rangle]^{;\nu} = 0. \label{second_term_divergence}
\end{align}

Summing Eqs.(\ref{first_term_divergence}) and (\ref{second_term_divergence}), one concludes that:
\begin{align}
    {T^{GW}_{\mu \nu}}^{;\nu} \neq 0. \label{vanishing_divergence}
\end{align}
This means that even in theories where scalar and tensorial waves are luminal and where the TT gauge is achievable, the GW EMT can still have non-vanishing divergence.

In \cite{Lobato2022}, we have shown how, in the cosmological context, the vanishing of the GW EMT divergence can be used to obtain the GW version of the duality relation \footnote{This relation was originally obtained by other means in \cite{Tasinato2021}.}:
\begin{align}
    D_{L}^{(GW)} = (1+z)^2 D_{A}^{(GW)}, \label{duality_relation}
\end{align}
where $z$ is the redshift, $D_L^{(GW)}$ is the GW luminosity distance and $D_A^{(GW)}$ the correspondent angular-diameter distance.
This relation, in turn, influence certain observables. The signal of a stochastic gravitational wave background is related to the spectral GW energy density
\begin{align}
    \langle S \rangle = \mathcal{F}\bigg(\frac{d^3 \rho_{gw}}{df d^2\Omega} \bigg),
\end{align}
where $f$ is the frequency and $\Omega$ is the solid angle of the observation. It is possible to obtain the total spectral energy density as a summation of the different galaxy sources contributions \cite{Cusin2017}:
\begin{align}
 	\frac{d^3 \rho_{GW}}{dfd^2\Omega} &=  \int \int \frac{1+z(\vartheta)}{4\pi} \left(\frac{D_A^{(GW)}}{D_L^{(GW)}}\right)^2 \sqrt{p^{\mu}(\vartheta)p_{\mu}(\vartheta)}\times\nonumber \\ &\hspace{30pt} \times n_G(x^{\mu}(\vartheta), \theta_G)\mathcal{L}_G(f_G, \theta_G) d\vartheta d\theta_G,  \label{SGWB_energy}
\end{align}
where $p_{\mu}$ is the spatial projection of the tangent vector of the GW null geodesics $x^{\mu}(\vartheta)$, $n_G$ is the number density of galaxies, $\mathcal{L}_G$ are the GW luminosity of the galaxies and $\theta_G$ are a set of variables related to the galaxy emitting the waves. 

We have seen that in more general reduced Horndeski theories, the GW EMT has non-vanishing divergence, which would imply in a modification to the duality relation and a consequent change on the stochastic GW signal, by Eq.~(\ref{SGWB_energy}). The stochastic GW background, in principle, can then serve as a test  reduced Horndeski theory. If any deviation from the expected signal predicted in the GR theory ever arise, this can indicate that more general theories must be considered to explain the phenomenon. 

\section{Conclusion}

The main results of this work are Eqs.~(\ref{complete_EMT}), (\ref{T_2}), (\ref{T_3}), (\ref{T_4}), with the subsequent coefficient definitions. It represents the off-shell GW EMT for the family of reduced Horndeski theories, with scalar waves included. With it, we were able to study some subcases of interest: those where scalar waves are luminal, Eq.(\ref{luminal_GW_EMT}), and those where the TT gauge is achievable, Eq.~(\ref{TT_EMT}). In the first of these, the trace of the tensor was shown to vanish, as expected for radiation fluids and in the second it was shown not to vanish trivially, Eq.~(\ref{TT_gauge}), without the imposition of the full field equations. As suggested by studies of the GW EMT in $f(R)$ theories \cite{Preston2014, Preston2016}, in the cosmological context, this non-vanishing of the trace could indicate a candidate to dark energy: GW curving background space-time could be part of the reason behind the current accelerated expansion era. 

The Brans-Dicke GW EMT was reobtained as a particular case, confirming previous results from literature \cite{Isi2018}. Different from previous literature, we obtain the GW EMT in Brans-Dicke for both the Jordan and Einstein frames, Eqs.~(\ref{Jordan_EMT}) and (\ref{Einstein_EMT}), respectively, and compare them to conclude that the tensor, and consequently the energy density of GWs, is not invariant under conformal transformations (i.e. Eqs.~(\ref{Jordan_to_Einstein_EMT}) and (\ref{Jordan_to_Einstein_energy})).

Furthermore, with the help of previous studies regarding GW propagation in reduced Horndeski theories \cite{Dalang2020, Dalang2021}, we can expect that, within particular theories where the divergence of the GW EMT does not vanish in the geometrical optics limit, the GW duality relation will be modified and, therefore, we get a first glimpse of how our results can entail changes in the SGWB observations, even after the imposition of the field equations: either (i) by directly modifying the GW spectral energy density per solid angle calculated in \cite{Cusin2017} as Eq.~(\ref{SGWB_energy}), in any background space-time,  or (ii) by the change in the relation between the SGWB detected signal and spectral energy density, as exemplified in particular theories by \cite{Isi2018}. We expect that different theories will predict different SGWB signals for a fixed set of GW sources and that this can be a way of probing deviations from the usual GR theory.

Our work complements the study made by \cite{Stein2011} in the sense that it investigates the GW EMT in a different set of scalar-tensor theories. In another sense, it generalizes this same work by allowing any background space-time, not only asymptotically-flat ones, which would make important cases such as FLRW universes to be left aside. 

Although full generality has been achieved in terms of the background metric, we must alert for the limits of our approach. The weak-limit averaging process used to compute the GW EMT can result in divergent terms in theories where $\bar{G}_{3,X} \neq 0$. Those potential divergences are all present in the part of the GW EMT given by Eq.~(\ref{T_3}). We have pointed out that such divergences can still vanish in the most troubling cases after imposing the field equations. When, even after this impositions,such terms remain, we point out that the problem can be in our assumption that the GW amplitude and wavelength are of the same order of smallness, as in Eq.~(\ref{amplitude_wavelength}). A more detailed treatment where different limiting procedures are done for both perturbative parameters might eliminate such divergences. 

Future steps could be made in the direction of generalizing the GW EMT for Horndeski theories in which tensorial perturbations are not luminal and for theories beyond Horndeski. Further studies on the relevance of our results to observations of the SGWB signal in Horndeski theories are worth being pursued as well. 

\section{Acknowledgments}

J.C.L. thanks Brazilian funding agency FAPERJ for PhD scholarship 2016007634.
\appendix

\section{Details regarding variation of action}

\label{appendix}

Here we provide details on how to obtain each contribution of Eq.~(\ref{T_second}). As discussed in the main text, the terms $T^{(i)}_{\mu \nu}$ are obtained by varying the contributions inside the average of Eq.~(\ref{T_first}) that are directly related with the functions $\bar{G}_i$. Using Eq.~(\ref{var_metric}), (\ref{var_X}) and (\ref{var_G}), one obtains:
\begin{widetext}
\begin{align}
T^{(second,2)}_{\mu \nu} = \frac{1}{16\pi G}\bigg\langle  &\epsilon^2\bigg\{\bar{G}_{2,X} \delta \varphi_{,\mu}\delta \varphi_{,\nu} + \bar{G}_{2,XXX}(\bar{ \varphi}_{,\alpha} \delta \varphi^{,\alpha})^2 \frac{\bar{ \varphi}_{,\mu} \bar{ \varphi}_{,\nu}}{2}-2\bar{G}_{2,XX}\bigg[\bar{ \varphi}_{(,\mu}\delta \varphi_{,\nu)} \bar{ \varphi}_{,\lambda} + \delta \varphi_{,\lambda}\frac{\bar{ \varphi}_{,\mu}\bar{ \varphi}_{,\nu}}{4}\bigg]\delta \varphi^{,\lambda}\bigg\} \bigg \rangle. 
\end{align}
\end{widetext}

The calculation of the remaining terms is a little more extensive. For the ones involving $\bar{G}_{4}$, for example, using the results of Subsec.~\ref{subsec:Variation_of covariant_derivatives}, one concludes that:
\begin{widetext}
\begin{align}
	T^{(second,4)}_{\mu \nu} = \frac{-1}{16\pi G} \bigg \langle  & \epsilon^2\frac{\bar{G}_4}{2} \bigg\{2\bigg(\frac{\delta h}{\delta \bar{g}^{\mu \nu}}\bigg)_{;\beta}h^{,\beta} + h_{,\mu} h_{,\nu} + 2\bigg(2\bigg(\frac{\delta h^{\alpha \gamma}}{\delta \bar{g}^{\mu \nu}}\bigg)^{;\beta} - \bigg(\frac{\delta h^{\alpha \beta}}{\delta \bar{g}^{\mu \nu}}\bigg)^{;\gamma}\bigg)h_{\alpha \beta;\gamma} -2 {h_{\mu \gamma}}^{;\beta}{h_{\nu \beta}}^{;\gamma} - h_{\alpha \beta;\mu}{h^{\alpha \beta}}_{;\nu} \nonumber \\ &\hspace{5pt}+ 2 h_{\mu \beta;\gamma}{h_{\nu}}^{\beta;\gamma} -2h_{,\beta}\bigg(\frac{\delta h^{ \beta \gamma}}{\delta \bar{g}^{\mu \nu}}\bigg)_{;\gamma}   -2\bigg(\frac{\delta h}{\delta \bar{g}^{\mu \nu}}\bigg)^{;\beta}{h_{\beta \gamma}}^{;\gamma}    \bigg \} + 2\epsilon^2 \bar{G}_{4,\varphi} \delta \varphi \bigg[\bigg(\frac{\delta h^{\alpha \beta}}{\delta \bar{g}^{\mu \nu}}\bigg)_{;\alpha \beta} - \bar{\Box}\bigg(\frac{\delta h}{\delta \bar{g}^{\mu \nu}}\bigg) \nonumber \\& \hspace{370pt}- h_{;\mu \nu}\bigg] \bigg \rangle \label{prel_T_4}
\end{align}
\end{widetext}
Using integration by parts (in the sense of Eq.(\ref{by_parts})) in terms where derivatives of perturbations are contracted, one will end up with contributions where $\bar{\Box} \hat{\gamma}_{\alpha \beta}$ appears. For example, using Eqs.(\ref{var_h}):
\begin{align}
\bar{G}_4\bigg\langle \epsilon^2\bigg(\frac{\delta h}{\delta \bar{g}^{\mu \nu}}\bigg)_{;\beta}h^{,\beta} \bigg\rangle &= -\bar{G}_4\bigg\langle2\epsilon^2h\bar{\Box}\bigg(\frac{\delta h}{\delta \bar{g}^{\mu \nu}}\bigg) \bigg\rangle \nonumber \\ &= \bar{G}_4\bigg\langle2\epsilon^2h\bigg[\bar{\Box}\hat{\gamma}_{\mu \nu}-\bar{F}_{\mu \nu}\bar{\Box}\delta \varphi\bigg] \bigg\rangle.
\end{align}
Since we are assuming luminal tensorial modes from the start, the equation for the tensorial part of GWs is of the form \cite{Dalang2021}:
\begin{align}
    \epsilon \bar{\Box} \hat{\gamma}_{\mu \nu} + \mathcal{O}(1) = 0, \label{luminal_tensorial_modes}
\end{align}
where $\epsilon \bar{\Box} \hat{\gamma}^{\alpha \beta} = \mathcal{O}(\epsilon^{-1})$. The terms involving $\bar{\Box} \hat{\gamma}^{\alpha \beta}$ appearing in the GW EMT are of the form $
\langle \epsilon^2 W \bar{\Box} \hat{\gamma}^{\alpha \beta} \rangle$, with $W = \mathcal{O}(1)$. Multiplying Eq.(\ref{luminal_tensorial_modes}) by $\epsilon W$ and taking the weak limit average:
\begin{align}
\langle \epsilon^2 W \bar{\Box} \hat{\gamma}^{\alpha \beta} \rangle = 0, \label{dalambertian_vanish_tensor}
\end{align}
so that these D'lembertians terms can be neglected. Of course, the same will be true for terms with $\bar{\Box} \hat{\gamma}$. Notice, however, that we are still interested in the case of non-luminal scalar waves, so that analogous terms involving $\bar{\Box} \delta \varphi$ do not vanish in principle.

Another simplification comes from using integration by parts in terms where divergences of $\hat{\gamma}_{\mu \nu}$ can appear. Because of the harmonic gauge given by Eq.~(\ref{harmonic_gauge}), these divergences vanish. For example, using Eqs.(\ref{h_in_terms_of_gamma}) and (\ref{derivative_exchange}):
\begin{align}
    \langle-2\epsilon^2{h_{\mu\gamma}}^{;\beta}{h_{\nu\beta}}^{;\gamma}\rangle &= \langle-2\epsilon^2{h_{\mu\gamma}}^{;\gamma}{h_{\nu\beta}}^{;\beta}\rangle \nonumber \\ &=   \bigg \langle\epsilon^2\bigg\{-\frac{\hat{\gamma}_{,\mu}\hat{\gamma}_{,\nu}}{2}  - 2 C_{ \gamma (\mu}\hat{\gamma}_{,\nu)} \delta \varphi^{,\gamma}  \nonumber \\ & \hspace{50pt}- 2 C_{\mu \gamma} C_{\nu \beta} \delta \varphi^{,\gamma} \delta \varphi^{,\beta} \bigg\}\bigg\rangle,
\end{align}
where the harmonic gauge was already imposed.

With these simplifications, expressing everything in Eq. (\ref{prel_T_4}) in terms of $\hat{\gamma}_{\mu \nu}$ and $\delta \varphi$ by using Eqs.(\ref{h_in_terms_of_gamma}), (\ref{var_h}) and (\ref{delta_h}) and adding the terms present in $T^{(first)}_{\mu \nu}$ (given by Eq.~(\ref{T_first})) that have the $\bar{G}_4$ function and its derivatives as factors, one is able to obtain 
Eq.(\ref{T_4}).

The last part of the GW EMT is $T^{(second,3)}_{\mu \nu}$. Different from the previous parts, one needs to take care when varying covariant derivatives here, as discussed in Subsec. \ref{subsec:Variation_of covariant_derivatives}. Variation will still commute with covariant derivatives when the field being differentiated is a perturbation. But in terms where derivatives act upon the background scalar (\textit{i.e.} Eq.~(\ref{var_background_scalar})), this cannot be done. There are two terms where this occurs, namely those with coefficients given by Eq.~(\ref{first_example}) and (\ref{second_example}). Initially, notice that
\begin{align}
    \delta (\bar{\varphi}_{;\mu \nu}) = - \bar{\varphi}_{,\tau} \delta \bar{\Gamma}^{\tau}_{\mu \nu}. 
\end{align}
It is straightforward to show that
\begin{align}
    \delta \bar{\Gamma}^{\tau}_{\mu \nu} = -\frac{1}{2}[\bar{g}_{\lambda \mu}(\delta \bar{g}^{\lambda \tau})_{;\nu}+\bar{g}_{\lambda \nu}(\delta \bar{g}^{\lambda \tau})_{;\mu} - \bar{g}_{\mu\gamma} \bar{g}_{\nu \sigma}(\delta \bar{g}^{\gamma \sigma})^{;\tau}]
\end{align}
Then, Eq.~(\ref{var_background_scalar}) becomes, after integration by parts and using that $L_{\alpha \beta}$ is a symmetric tensor in both cases of interest:
\begin{align}
    T_2 = -\frac{\epsilon^2}{2} \int \sqrt{-\bar{g}} [ 2({L_{\lambda}}^{ \beta}\bar{\varphi}_{,\sigma})_{;\beta}-(L_{\lambda \sigma}\bar{\varphi}_{, \gamma})^{;\gamma} ] \delta \bar{g}^{\lambda \sigma} d^4x, 
\end{align}
from which we conclude that the contribution for the GW EMT coming from the variation of connections are:
\begin{align}
    T_{\mu \nu} \sim \langle \epsilon^2[2({L^{ \beta}}_{(\mu}\bar{\varphi}_{,\nu)})_{;\beta}-(L_{\mu \nu}\bar{\varphi}_{, \gamma})^{;\gamma}] \rangle.
\end{align}
For the case in which Eq.~(\ref{first_example}) is valid:
\begin{align}
     T_{\mu \nu} \sim -\langle \epsilon^2 [&\bar{G}_{3,X} \delta \varphi_{,\gamma} \delta \varphi^{,\gamma} \bar{\varphi}_{(,\mu}]_{;\nu)} \rangle \nonumber \\ &\hspace{10pt}+ \frac{\bar{g}_{\mu \nu}}{2}\langle  \epsilon^2[\bar{G}_{3,X} \delta \varphi_{,\gamma} \delta \varphi^{,\gamma} \bar{\varphi}_{,\phi}]^{;\phi}\rangle.
\end{align}
For the case of Eq.~(\ref{second_example}):
\begin{align}
T_{\mu \nu} \sim \langle \epsilon^2 [&\bar{G}_{3,XX} (\delta \varphi_{,\gamma} \bar{ \varphi}^{,\gamma})^2 \bar{\varphi}_{(,\mu}]_{;\nu)} \rangle \nonumber \\ &\hspace{10pt}- \frac{\bar{g}_{\mu \nu}}{2}\langle  \epsilon^2[\bar{G}_{3,XX} (\delta \varphi_{,\gamma} \bar{ \varphi}^{,\gamma})^2 \bar{\varphi}_{,\phi}]^{;\phi}\rangle.
\end{align}
We confirm that these terms will not vanish in the weak limit average, different from the variations of connections coming from covariant derivatives of perturbations. 

With these terms in mind and varying the rest of the $\bar{G}_3$ related terms of Eq.~(\ref{T_first}), one obtains:
\begin{widetext}
	\begin{align}
		T^{(second,3)}_{\mu \nu}=\frac{1}{16\pi G} \bigg \langle & \epsilon^2 \bigg\{ \bar{G}_{3,\varphi X}\bar{ \varphi}_{,\mu}\bar{ \varphi}_{,\nu}\delta \varphi \bar{\Box}\delta \varphi-2\bar{G}_{3,\varphi}\delta \varphi \delta \varphi_{;\mu \nu} +\bar{G}_{3,XXX}\frac{\bar{\varphi}_{,\mu}\bar{\varphi}_{,\nu}}{2}\bar{\Box}\bar{ \varphi}(\bar{ \varphi}_{,\gamma}\delta \varphi^{,\gamma})^2 + \frac{\bar{g}_{\mu \nu}}{2}[\bar{G}_{3,X}\delta \varphi_{,\gamma}\delta \varphi^{,\gamma}\bar{ \varphi}_{,\delta}]^{;\delta} \nonumber \\ & - [\bar{G}_{3,X}\delta \varphi_{,\gamma}\delta \varphi^{,\gamma}\bar{ \varphi}_{,(\mu}]_{;\nu)}+[\bar{G}_{3,XX}(\bar{ \varphi}_{,\gamma}\delta \varphi^{,\gamma})^2\bar{\varphi}_{(,\mu}]_{;\nu)}-\frac{\bar{g}_{\mu \nu}}{2}[\bar{G}_{3,XX}(\bar{ \varphi}_{,\gamma}\delta \varphi^{,\gamma})^2\bar{\varphi}_{\beta}]^{;\beta}\nonumber \\ &- \bar{G}_{3,X}\bigg\{\bar{ \varphi}_{,\alpha}\bar{ \varphi}_{,\beta}\bigg[\frac{\delta h^{\alpha \beta}}{\delta \bar{g}^{\mu \nu}}\bar{\Box}\delta \varphi + h^{\alpha \beta}\delta \varphi_{;\mu \nu} -\delta \varphi^{,\alpha}\bigg[\bigg(\frac{\delta h}{\delta \bar{g}^{\mu \nu}}\bigg)^{;\beta} -2\bigg(\frac{\delta h^{\beta \gamma}}{\delta \bar{g}^{\mu \nu}}\bigg)_{;\gamma}\bigg]\bigg]\nonumber \\ &  \hspace{40pt}-\bar{ \varphi}_{,\beta}\bar{ \varphi}_{(,\mu}\delta \varphi_{,\nu)}(h^{,\beta} - 2{h^{\beta \gamma}}_{;\gamma}) - \delta \varphi^{,\alpha} h_{(,\mu}\bar{ \varphi}_{,\nu)} \bar{ \varphi}_{,\alpha} - \delta \varphi_{,\mu} \delta \varphi_{,\nu} \bar{\Box}\bar{ \varphi} -  \delta \varphi_{,\alpha} \delta \varphi^{,\alpha} \bar{ \varphi}_{;\mu \nu}\nonumber \\ & \hspace{245pt}- \bar{ \varphi}_{,(\mu}\delta \varphi_{,\nu)}\bar{\Box}\delta \varphi - \bar{ \varphi}_{,\alpha} \delta \varphi^{,\alpha}\delta \varphi_{;\mu \nu}\bigg\} \nonumber \\ &  -2\bar{G}_{3,XX}\bigg\{\frac{(\bar{\varphi}_{,\gamma}\delta \varphi^{,\gamma})^2}{2}\bar{\varphi}_{;\mu \nu}-\frac{\bar{ \varphi}_{,\mu}\bar{ \varphi}_{,\nu}}{4}\{\bar{ \varphi}_{,\alpha}\bar{ \varphi}_{,\beta}[h^{\alpha \beta}\bar{\Box}\delta \varphi -\delta \varphi^{,\alpha}(h^{,\beta}- 2{h^{ \beta \gamma}}_{;\gamma})] - \delta \varphi_{,\alpha} \delta \varphi^{,\alpha}\bar{\Box}\bar{ \varphi}\nonumber \\ & \hspace{210pt} - \delta \varphi_{,\alpha} \bar{ \varphi}^{,\alpha} \bar{\Box}\delta \varphi\} + \bar{ \varphi}_{,\gamma}\delta \varphi^{,\gamma}\bar{ \varphi}_{(,\mu}\delta\varphi_{,\nu)}\bar{\Box}\bar{ \varphi}\bigg\} \bigg\} \bigg\rangle, \label{T_second_3}
	\end{align}
\end{widetext}
By expressing the above equation in terms of $\hat{\gamma}_{\mu \nu}$ and $\delta \varphi$ and adding the terms present in $T^{(first)}_{\mu \nu}$ (given by Eq.~(\ref{T_first})) having the $\bar{G}_3$ function and its derivatives as factors, one is able to obtain 
Eq.(\ref{T_3}).


\begin{thebibliography}{99}
    
    \bibitem{Agazie2023I} G. Agazie et al., ApJL 951 L8 (2023).

    \bibitem{Agazie2023II} G. Agazie et al., ApJL 952 L37 (2023).

    \bibitem{Afzal2023} A. Afzal et al., ApJL 951 L11 (2023).
    \bibitem{Amaro2017} P.Amaro-Seoane et al., arXiv:1702.00786 (2017).
    \bibitem{Baker2020} J.Baker et al., arXiv:1907.06482 (2020).
    \bibitem{Barausse2020} E. Barausse, Gen. Relativ. and Grav. 52:81 (2020).

    \bibitem{Amaro2023}  P.Amaro-Seoane et al., Liv. Rev. Relativ. 26, 2 (2023).
    \bibitem{Auclair2023} P. Auclair et al., Liv. Rev. Relativ. 26, 5 (2023).
    
    \bibitem{Romano2017} J. Romano and N. Cornish, Liv. Rev. Relativ. 20, 2 (2017).

    \bibitem{Abbot2018} B. P. Abbot et al. (Virgo and LIGO Scientific Collaborations), Phys. Rev. Lett. 120, 201102 (2018).

    \bibitem{Isi2018} M. Isi and L. C. Stein, Phys. Rev. D 98, 104025 (2018).
    
    \bibitem{Lobato2022} J. C. Lobato, I. S. Matos, M. O. Calv\~ao and I. Waga, Phys. Rev. D 106, 104048 (2022). 

    \bibitem{Abbot2017} B. P. Abbott et al.(Virgo and LIGO Scientific Collaborations),Phys. Rev. Lett. 119, 161101 (2017).

    \bibitem{Goldstein2017} A. Goldstein et al., Astrophys. J. 848, L14 (2017).
    \bibitem{Isaacson1968I} R. A. Isaacson, Phys. Rev. 166, 1263 (1968).
    \bibitem{Isaacson1968II} R. A. Isaacson, Phys. Rev. 166, 1272 (1968).

    \bibitem{Green2011} S. R. Green and R. M. Wald, Phys. Rev. D 83, 084020 (2011).

    \bibitem{Green2013} S. R. Green and R. M. Wald, Phys. Rev. D 87, 124037 (2013).
    \bibitem{Burnett1989} G. A. Burnett, J. Math. Phys. 30, 90 (1989).
    \bibitem{Stein2011} L. C. Stein and N. Yunes, Phys. Rev. D 83, 064038 (2011).

    \bibitem{Preston2014} A. W. H. Preston, T. R. Morris, JCAP 09 (2014) 017.
    \bibitem{Preston2016} A. W. H. Preston, JCAP 08 (2016) 038.
    \bibitem{Dalang2020} C. Dalang, P. Fleury and L.Lombriser, Phys. Rev. D 102, 044036 (2020).	

    \bibitem{Dalang2021} C. Dalang, P. Fleury and L. Lombriser, Phys. Rev. D 103, 064075 (2021).
    
    \bibitem{Kobayashi2019} T. Kobayashi, Rep. Prog. Phys (2019).

    \bibitem{Garoffolo2019} A. Garoffolo, G. Tasinato, C. Carbone, D. Bertacca, and S. Matarrese, arXiv:1912.08093.

    \bibitem{Gravitation} C. W. Misner, K. S. Thorne, and J. A. Wheeler, ``Gravitation" (W. H. Freeman and Company, San Francisco, 1973), pp. 964.
    
    \bibitem{MaggioreVol1} M. Maggiore, Gravitational Waves, Vol. 1: Theory and Experiments (Oxford University Press, New York, 2007).

    \bibitem{Dinverno}R. D'inverno, Introducing Einstein's relativity (Oxford University Press, Oxford, England, 1992). 
    \bibitem{Tasinato2021} G. Tasinato et al., JCAP 06 (2021) 050.
    
\bibitem{Capozziello2011} S. Capozziello and V. Faraoni, Beyond Einstein gravity: A survey of gravitational theories for cosmology and astrophysics (Springer, New York, United States, 2011). 

  \bibitem{Cusin2017} G. Cusin, C. Pitrou and J.-P. Uzan, Phys. Rev. D 96, 103019 (2017).
\end{thebibliography}
\end{document}